\documentclass[reprint,onecolumn,eqsecnum,amsmath,nofootinbib]{revtex4-2}

\usepackage{graphicx}

\usepackage[colorlinks=true, allcolors=black]{hyperref}

\begin{document}

\title{Building post-Newtonian neutron stars}

\author{Nils Andersson, Fabian Gittins,  Shanshan Yin and Rodrigo Panosso Macedo}

\affiliation{
Mathematical Sciences and STAG Research Centre, University of Southampton,
Southampton SO17 1BJ, United Kingdom}

\begin{abstract}
Owed to their compactness, neutron stars involve strong gravity and extreme density physics. Nevertheless, at present, there are a variety of problems where progress (at least conceptually) can be made in the context of weak gravity. Motivated by this we examine how accurately one can model neutron stars using the post-Newtonian approximation to general relativity. In general, we find there is a significant degree of freedom in how the post-Newtonian equations of stellar structure can be formulated. We discuss this flexibility in the formulation and provide examples to demonstrate the impact on stellar models. We also consider the (closely related) problem of building neutron stars using isotropic coordinates. In this context, we provide a new strategy for solving the equations (based on a scaling argument) which significantly simplifies the problem.
\end{abstract}

\maketitle

\section{Introduction}

Neutron stars involve extremes of physics---in terms of density, magnetic fields, temperature and so on---that make them interesting from many different perspectives. Their special place in the pantheon of  astrophysics, nuclear and gravitational physics is justified by a rich observational phenomenology and a range of outstanding issues that remain to be explained/understood \cite{NAbook}. The recent focus on neutron stars as sources of gravitational waves, both in binary systems (like the celebrated case of GW170817 \cite{2017PhRvL.119p1101A}) and as emitters of continuous waves \cite{2022arXiv220606447R}, bring a number of modelling questions into play. The specific technical challenges may differ but the overarching question remains the same: how do we appropriately represent the physics required to describe a given scenario? By necessity, this has become a gradual grind towards (hopefully) eventual enlightenment.

With the many difficult questions in mind, the scope of the present discussion may seem quite modest. We want to understand how accurately---or not---we can describe neutron stars within the post-Newtonian (pN) approximation (formally, a weak-field, slow-motion expansion of general relativity) \cite{PWbook}. At first sight, this may seems like a somewhat misguided exercise. After all, due to their large mass $M$ and small radius $R$, neutron stars are firmly beyond the pN regime. With a compactness of order
\begin{equation}
    \mathcal C = {GM \over Rc^2} \approx 0.2
\end{equation}
(where $G$ is Newton's gravitational constant and $c$ is the speed of light) it is not at all clear that second order pN terms (of order $1/c^4$ in a formal expansion) can be neglected compared to first order ones ($1/c^2$). This is, of course, well understood and a good reason to use a fully relativistic approach to studies of neutron-star dynamics. Having said that, one may still be motivated to consider the pN problem. First and foremost, it is important to establish how the star responds to the presence of a binary partner and what this implies for the standard pN expansion for the gravitational-wave signal from an inspiralling binary. This problem is typically expressed as an expansion for small orbital frequencies \cite{PWbook}, while the star's response to the tidal interaction is more immediately connected to $\mathcal C$ and/or the characteristic frequencies of neutron-star dynamics. This ``mismatch'' is well illustrated by the role of the (static) tidal deformability, which formally enters at a high (orbital) pN order but which nevertheless is expected to be detectable with advanced instruments \cite{2022PhRvD.105h4021C}. A closely related issue that springs to mind is the mode-sum approach to the dynamical tides in neutron-star binaries, see \cite{2020PhRvD.101h3001A} for a recent discussion. In the Newtonian case, the solution to this problem relies on the orthogonality and  completeness of the star's oscillation modes \cite{1978ApJ...221..937F,1978ApJ...222..281F}. Both  properties are complicated in general relativity---in fact, the modes are not going to be complete due to the existence of the late-time power-law tail in the response to a perturbing agent \cite{1994PhRvD..49..883G,1999CQGra..16R.159N}---but will remain so in the pN description (at least up to, and including, 2pN terms, which is further than we intend to proceed here). The mathematics will be  messier than in the Newtonian case, as exemplified by \cite{1965ApJ...142.1488C}, but so be it. The main question becomes one of precision---are pN model ``accurate'' enough to be useful for quantitative models of (say) neutron-star tides? This is not a new question (see, \textit{e.g.}, \cite{2015PhRvD..92b4056G}, although their focus on alternative theories of gravity is rather different from what we consider here) but we hope to demonstrate that it is worth revisiting. Immediate motivation is provided by the alternative pN descriptions in the literature---from Chandrasekhar's classic work on pN hydrodynamics (which strictly truncates the equations at 1pN order) \cite{1965ApJ...142.1488C} to the modern description in the  monograph from Poisson and Will (which, as we will discuss, retains some 2pN terms in the first order equations) \cite{PWbook} and the formulation of Blanchet, Damour and Schaefer which aims to stay close to the variables used in numerical relativity (and which also mixes in 2pN terms) \cite{1990MNRAS.242..289B}.

If we assume that any  2pN terms may be included without penalty (which is true in the regime where the pN approach truly applies) then the different formulations are all equivalent. However, for actual neutron-star models the inclusion of 2pN terms  makes a numerical difference. Hence, it is worth asking if one of the pN prescriptions fares (numerically) better than the others when we employ a realistic matter description, or are they all equally ``bad''? This is the question we explore in this paper. It may not seem particularly interesting at first, but a number of subtle issues come into play, making the detailed analysis worthwhile.

The paper is organised as follows. We begin in section~\ref{sec:build} with a summary of the traditional approaches to solving the stellar structure equations in Newtonian and relativistic gravity. This motivates the truncated form of the pN equations. In section~\ref{sec:isotropic}, we formulate the fully relativistic problem (and its corresponding truncated pN form) in isotropic coordinates. This may be a brief aside, but it is important as the pN formulation is commonly based on the use of isotropic coordinates. In section~\ref{sec:formulating}, following the standard pN formulation of fluid dynamics, we discuss the (perhaps surprising) flexibility that is afforded in pN theory. We demonstrate how this flexbility impacts on constructed neutron-star models. Finally, we conclude in section~\ref{sec:summary}.

\section{How to build a star}
\label{sec:build}

The issue we want to explore involves solving the equations of hydrostatic equilibrium for a self-gravitating body. This is (obviously) a textbook problem  so it might be tempting to  bypass the introductory aspects and cut straight to the chase. However, in order to understand the issues that arise later it is helpful to start with a few comments on what may be very familiar results. 

\subsection{The Newtonian case}

Not wanting to miss an opportunity to set the bar as low as possible, let us start with the Newtonian gravity problem. Then we have the mass density $\rho$, which leads to the mass $M_N$ through
\begin{equation}
  \frac{dM_N}{d\xi} = 4 \pi \xi^2 \rho \ . 
\end{equation}
For reasons that will become clear later, we are using $\xi$ to represent the radial coordinate here. We also need the pressure $p$, which in equilibrium follows from
\begin{equation}
 \frac{dp}{d\xi} =
    -{G\rho M_N \over \xi^2}  
\end{equation}
and the gravitational potential $U_N$, obtained from
\begin{equation}
     \frac{dU_N}{d\xi} = - \frac{G M_N}{\xi^2} \ .
\end{equation}
These three equations form a system that, when complemented by a matter equation of state $p=p(\rho)$, can be solved to obtain a stellar configuration. The surface of the star---at radius $\xi=R_s$---is simply obtained from $p(R_s)=0$. The salient point (for future reference) is that we can solve for the mass and radius of a star without involving the gravitational potential: the equation for $U_N$ decouples from the other two. This is convenient because the potential involves the a priori unknown value at the centre of the star, $U_0=U_N(\xi=0)$. After integration, the value of $U_0$ is fixed by matching the solution to the external potential 
\begin{equation}
    U_N(R_s) = {GM_N(R_s) \over R_s}
\end{equation}
So far, so good. Unfortunately, Newtonian gravity does not provide an accurate representation of neutron stars. We need to use a realistic equation of state (obtained from nuclear physics) and it is well known that the associated internal energy has a significant impact on the star's mass and radius.  Given this, let us consider the (similarly textbook) problem in general relativity. 

\subsection{The TOV equations}

The equations that describe relativistic hydrostatic equilibrium are known as the Tolman-Oppenheimer-Volkoff (TOV) equations. They are typically expressed in terms of Schwarzschild coordinates associated with the (general spherically symmetric) line element 
\begin{equation}
    ds^2 = - e^\nu c^2 dt^2 + e^\lambda d\xi^2 + \xi^2 \left( d\theta^2 + \sin^2\theta d\varphi^2 \right) \ , 
\end{equation}
where we are (again) using $\xi$ to represent the (now circumferential) radial coordinate. The problem involves two metric potentials, $\nu$ and $\lambda$, both functions only of $\xi$. The standard approach is to introduce a mass $M_g$ such that
\begin{equation}
    {dM_g \over d\xi} = {4\pi \over c^2} \xi^2  \varepsilon \ ,
\end{equation}
where $\varepsilon$ is the total energy density. It then follows that
\begin{equation}
    e^{-\lambda} = 1 - {2GM_g \over \xi c^2}  \ , 
\end{equation}
which matches the Schwarzschild exterior if we identify $M_g(R_s)$ as the gravitational mass, in turn defined by the asymptotic behaviour of the spacetime metric.

The pressure now follows from
\begin{equation}
       \frac{dp}{d\xi} = - \frac{G}{c^2 \xi^2} ( \varepsilon + p) \left( M_g + \frac{4\pi \xi^3 p }{c^2}  \right) \left( 1 - \frac{2GM_g} {\xi c^2} \right)^{-1}  \ , 
\label{TOV}
\end{equation}
where the total energy can be written
\begin{equation}
    \varepsilon = \rho c^2 \left( 1 + {\Pi \over c^2}\right) \ , 
    \label{toten}
\end{equation}
with $\Pi$  the (specific) internal energy.
Finally, the remaining metric potential is determined by integrating
\begin{equation}
    {d\nu \over d\xi} = - {p+\varepsilon \over 2} {dp \over d\xi} \ .
\end{equation}

The pattern is familiar from the Newtonian case. The potential $\nu$ decouples from the equations for the mass and the pressure. We can solve the first two equations for the mass and radius of a star. This is useful because the third equation involves the initially unknown value of the potential at the centre, $\nu_0=\nu(\xi=0)$. This value has to (again) be determined by matching to the external gravitational field at the star's surface. 

The solution to the TOV equations is quite straightforward and it may seem self-evident that it should be so. However, there are subtleties here. For example, it is worth noting that 
\begin{equation}
    e^{-\lambda} \to 1 \quad \mbox{as} \quad \xi\to 0
\end{equation}
ensures that the solution is spatially flat at the star's centre. Also, it is not immediately obvious that that the mass function $M_g$ represents the gravitational mass: Intuitively, the total energy contained inside the given volume. In order to argue that this is, indeed, the case it is natural to take a detour and consider the Newtonian virial theorem.

\subsection{The (Newtonian) virial theorem}
\label{virel}

As will become evident later, one may want to consider different mass functions. One way to do this, certainly within the pN framework, is to make use of the standard virial theorem. 
Starting from the gravitational potential energy $\Omega$, we have
\begin{multline}
    \Omega = - \int_0^{R_s} {G\rho M_N(\xi) \over \xi} dV =  -4\pi G \int_0^{R_s} \rho M_N(\xi) \xi d\xi = 4\pi \int_0^{R_s} \rho {dU_N\over d\xi}  \xi^3 d\xi  \\
    = 4\pi \int_0^{R_s} {dp\over d\xi} \xi^3 d\xi  = \underbrace{ \left[ 4\pi p \xi^3 \right]_0^{R_s} }_{=0}
    - 3 \int_0^{R_s} 4\pi p\xi^2 d\xi = - 3\int_0^{R_s} p dV \ .
\end{multline}
Alternatively, we may use the Poisson equation 
\begin{equation}
{d\over d\xi} \left( \xi^2 {dU_N\over d\xi} \right) = \xi^2 {d^2 U_N \over d\xi^2} + 2\xi {dU_N\over d\xi} =  - 4\pi G \rho \xi^2
\end{equation}
to get (noting the limits of integration)
\begin{multline}
    \Omega = - {1\over G} \int_0^\infty \xi {dU_N\over d\xi} {d\over d\xi} \left( \xi^2 {dU_N\over d\xi} \right)  d\xi \\
    = - {1\over G} \underbrace{ \left[ \xi^3 \left( {dU_N\over d\xi}\right)^2 \right]_0^\infty  }_{=0} + {1\over G} \int_0^\infty \left[  \xi^2 \left( {dU_N\over d\xi} \right)^2 + \xi^3 {dU_N\over d\xi}  {d^2 U_N \over d\xi^2} \right] d\xi  \\ = - {1\over G} \int_0^\infty \xi^2  \left( {dU_N\over d\xi}\right)^2 d\xi - \int_0^\infty 4\pi  \rho \xi^3 {dU_N\over d\xi} d\xi
    =  - {1\over G} \int_0^\infty \xi^2  \left( {dU_N\over d\xi}\right)^2 d\xi - \Omega \ , 
\end{multline}
so 
\begin{multline}
    \Omega = - {1\over 2 G} \int_0^\infty  \xi^2  \left( {dU_N\over d\xi}\right)^2 d\xi = {1\over 2} \int_0^\infty M_N {dU_N\over d\xi} d\xi \\
    = {1\over 2} \underbrace{ [ M_N U_N]_0^\infty}_{=0} - {1\over 2} \int_0^{R_s} 4\pi  \rho \xi^2 U_N d\xi = - {1\over 2} \int_0^{R_s} \rho U_N dV \ .
\end{multline}
In summary, we have 
\begin{equation}
    \Omega = - \int_0^{R_s} {G\rho M_N \over \xi} dV = - 3\int_0^{R_s} p dV = - {1\over 2} \int_0^{R_s} \rho U_N dV \ .
\end{equation}
The key point is that, while the integrands may not be identical for a given value of $\xi$, when integrated over the star these relations should hold. 

We can use the virial argument to shed light on the mass used in the TOV equations (following an argument from Box 23.1 in \cite{2017grav.book.....M}). In Schwarzschild coordinates we have 
\begin{equation}
    M_g c^2 = 4\pi \int_0^{R_s} \varepsilon \xi^2 d\xi = \int_0^{R_s} \left( \varepsilon e^{-\lambda/2}\right) \underbrace{4\pi e^{\lambda/2} \xi^2 }_{=\sqrt{ ^{(3)}g}} d\xi = \int_0^{R_s} \varepsilon e^{-\lambda/2} dV \ .
\end{equation}
Making use of the virial relations---at the integrand level and discarding higher order pN terms---we have 
\begin{multline}
    \varepsilon e^{-\lambda/2} \approx \varepsilon \left( 1 - {GM_g(\xi) \over \xi c^2} \right)= c^2 \rho \left( 1 + {\Pi \over c^2} \right) \left( 1 - {GM_g(\xi) \over \xi c^2} \right) \\
    \approx c^2 \left[ \rho \left( 1 + {\Pi \over c^2} \right)  \underbrace{- \rho {GM_N(\xi) \over \xi c^2}}_{=\Omega/c^2} \right] =  c^2 \rho \left[ 1   + {1\over c^2} \left(  \Pi - { U_N \over 2}   \right)\right] \ .
\end{multline}
From this we identify the total energy as the sum of the mass, the internal energy and the gravitational potential energy. As these are the expected contributions to the total energy for a static star, it makes sense to conclude that the mass $M_g$ represents the gravitational mass.  The argument is, of course, only valid at 1pN order, but this is good enough for our purposes.

\subsection{Truncating the TOV equations} 
\label{tovtrunc}

\begin{figure}
    \centering
    \includegraphics[width=0.75\textwidth]{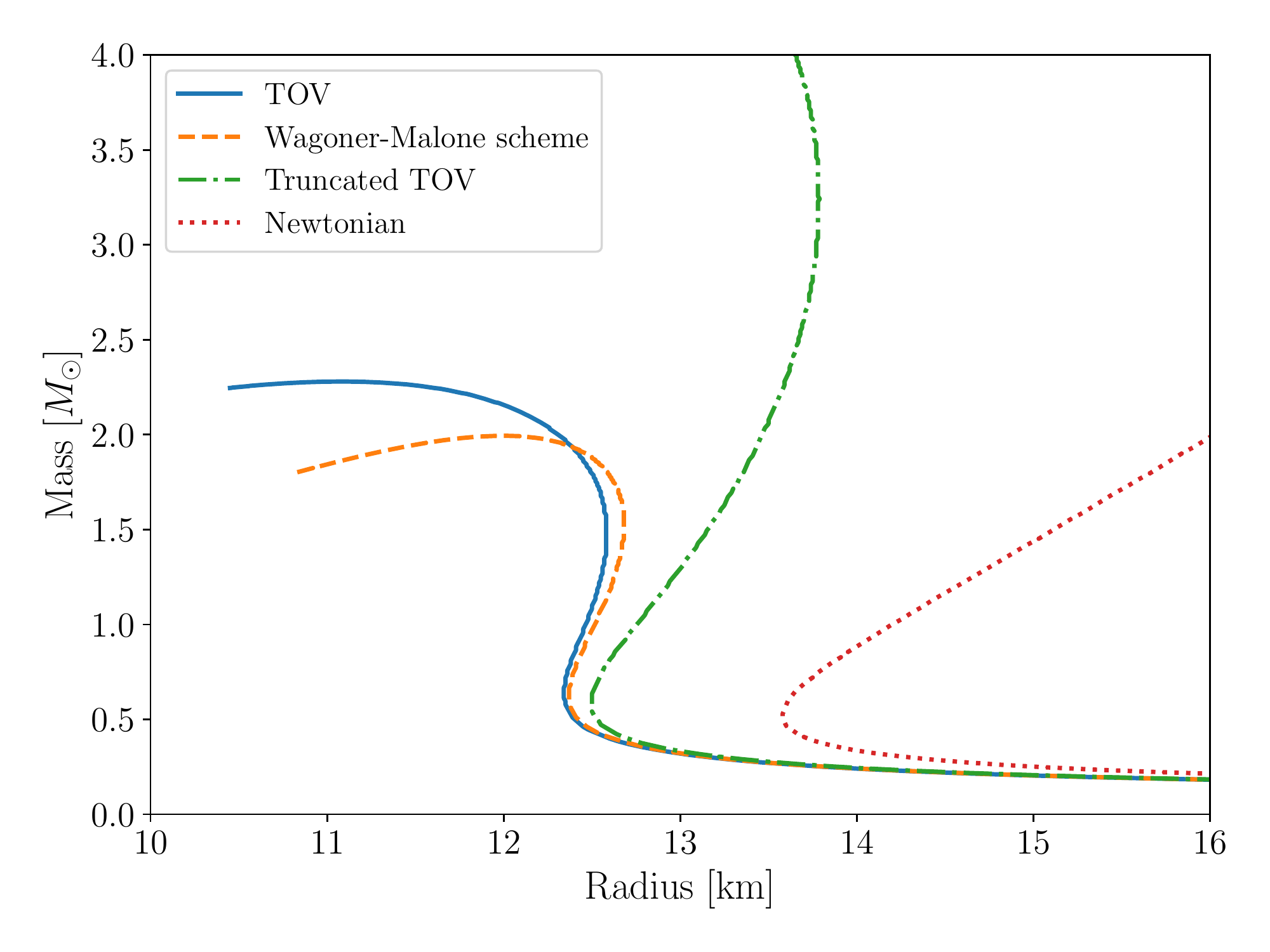}
    \caption{Comparison of stellar models obtained from the TOV equation \eqref{TOV} (blue solid), and the truncated equations \eqref{pboxed} (green dash-dot) and  \eqref{wagmal} (orange dashed). The data correspond to the BSk24 equation of state. For reference we also show the corresponding Newtonian result (red dotted).}
    \label{TOVfig}
\end{figure}

As we are interested in making contact with the pN expansion, we can truncate the TOV equations at some given order of $1/c^2$ \cite{1999PhRvD..60f7504S,2015PhRvD..92b4056G}. As a first step in this direction we may expand the right-hand side of \eqref{TOV} as 
\begin{equation}
   \frac{dp}{d\xi} 
    \approx  - \frac{G}{c^2 \xi^2}  (\varepsilon +p)  \left( M_g + \frac{4\pi \xi^3 p}{c^2}  \right) \left( 1 + \frac{2GM_g} {\xi c^2} \right)
    \approx 
    -{G M_g \over  \xi^2} { \varepsilon + p \over c^2} \left[ 1 + {1\over c^2} \left( \frac {4\pi \xi^3 p} {M_g} + \frac{2GM_g}{\xi} \right) \right] \ .
\end{equation}
This truncated TOV problem was explored by \citet{1999PhRvD..60f7504S}, who introduced $\varepsilon= \rho_t c^2$ to get 
\begin{equation}
   \frac{dp}{d\xi} \approx
    -{G M_g \rho_t \over  \xi^2} \left[ 1 + {1\over c^2} \left( {p\over \rho_t}+ \frac {4\pi \xi^3 p} {M_g} + \frac{2GM_g}{\xi} \right) \right] \ , 
\end{equation}
and then studied the convergence of the expansion as higher order $1/c^2$ terms were added in. We are not going to follow precisely this strategy because, in the pN approach it seems more natural to  identify the rest-mass density $\rho=m_B n$, where $n$ is the baryon number density\footnote{Keeping in mind that (cold) neutron-star equations of state tend to be provided as a table of $[n,\varepsilon,p]$.} and $m_B$ the baryon mass, and the internal energy density $\rho\Pi$ as in \eqref{toten}. 
This then leads to (see the post-TOV discussion by \citet{2015PhRvD..92b4056G})
\begin{equation}
    \frac{dp}{d\xi} \approx
    -{G\rho M_g \over \xi^2} \left[ 1 + \frac{1}{c^2}\left(\Pi + \frac{p}{\rho}+ \frac {4\pi \xi^3 p} {M_g} + \frac{2GM_g}{\xi} \right) \right] \ .
    \label{pboxed}
\end{equation}
Notably, this equation is not ``consistent'' in the pN sense as the mass $M_g$ itself contains order $1/c^2$ contributions, thereby introducing 2pN terms. As an illustration of the accuracy of the truncated model, consider the results in figure~\ref{TOVfig}, obtained for the BSk24 equation of state \cite{2013PhRvC..88b4308G}. Although we focus on a single equation of state for the nuclear matter---the same in all examples we provide---we have considered a wide range of matter models to confirm that the qualitative results remain the same. Evidently, the truncated model is good for stars below about $0.5M_\odot$ or so. As we increase the density the mass-radius curve diverges from the TOV result also demonstrating the absence of a maximum mass configuration. This is as expected for truncated/pN models. For comparison, we also show the mass-radius curve obtained from the Newtonian equations.  As an approximation, the Newtonian result clearly breaks down for even lower masses. This is as it should be. There are no real surprises here.

We have already touched upon the notion of different mass functions, hinting  that we may profitably work with new---perhaps less intuitive---variables. As an example,  let us  introduce  
\begin{equation}
    \bar m = M_g + {1\over c^2} \left( {GM_g^2 \over \xi} + 4\pi \xi^3 p \right) \ ,
\end{equation}
motivated by the right-hand side of \eqref{pboxed}. This then leads to a different truncated model
\begin{equation}
 \frac{dp}{d\xi} = - \frac{G \rho}{\xi^2} \bar{m} \left[ 1 + \frac{1}{c^2} \left( \Pi + \frac{p}{\rho} + \frac{G \bar{m}}{\xi} \right) \right]  \ ,
 \label{wagmal}
\end{equation}
where 
\begin{equation}
  \frac{d\bar{m}}{d\xi} = 4 \pi \xi^2 \rho \left[ 1 + \frac{1}{c^2} \left( \Pi + \frac{G \bar{m}}{\xi} - \frac{G \bar{m}^2}{4 \pi \xi^4 \rho} + 3 \frac{p}{\rho} \right) \right]  \ . 
\end{equation}
Incidentally, this form of the equations was first derived by \citet{1974ApJ...189L..75W} within the parameterised pN formalism (we will return to this derivation later). Clearly, these equations also include higher order pN terms, although they are not the same as in \eqref{pboxed}: the two models differ at order $1/c^4$. At the surface of the star, we now extract the gravitational mass from 
\begin{equation}
   M_g \approx \bar m  \left( 1 -  {G \bar m \over R_s c^2} \right) \ .
\end{equation}
Does this new mass function make a difference? Evidently so. The example provided in figure~\ref{TOVfig} shows that \eqref{wagmal}
provides much more accurate stellar models than the alternative truncation \eqref{pboxed}. This is interesting, but at the same time somewhat disturbing. If we can use the freedom to work with different mass functions, basically differing at the higher-order level, to our advantage then how do we know that we are making the optimal choice?

\section{Isotropic coordinates}
\label{sec:isotropic}

Turning to a different aspect of the standard pN formulations, but for the moment staying in the comfortable environment of general relativity, let us see what happens if we try to solve for hydrostatic equilibrium using isotropic coordinates (a problem also considered when constructing initial data for numerical relativity simulations, see \textit{e.g.} \cite{2015PhRvD..91j4030T}). One reason for doing this is that the results will later provide a direct comparison with the pN results (rather than relying on a ``translation'' from Schwarzschild coordinates). We will also find  the exercise instructive in its own right. 

\subsection{Decoupling the metric potentials}

In isotropic coordinates, the spatial part of the metric is taken to be conformally flat. That is, we have 
\begin{equation}
    ds^2 = - e^\nu c^2 dt^2 + e^\mu \left( dr^2 + r^2 d\theta ^2 + r^2 \sin^2 \theta d\varphi^2 \right) \ .
\end{equation}
Again, the gravitational mass is inferred from matching to the vacuum result. In the vacuum exterior, we now have
\begin{equation}
    ds^2 = - \left( 1- {GM_g/2rc^2} \over 1+ {GM_g/2rc^2} \right)^2 c^2 dt^2 + \left( 1+ GM_g/2rc^2 \right)^4 \left( dr^2 + r^2 d\theta ^2 + r^2 \sin^2 \theta d\varphi^2 \right) \ ,
    \label{isoSch}
\end{equation}
where $M_g$ is the  gravitational mass (as before) and 
we see that the transformation to Schwarzschild coordinates (with radial coordinate $\xi$, as before) is
\begin{equation}
    \xi = r \left( 1 + {GM_g \over 2rc^2} \right)^2 \ .
\end{equation}
In the weak field limit we get
\begin{equation}
    ds^2 \approx - \left( 1 - {2GM_g \over r c^2}  \right) c^2 dt^2 + \left( 1 + {2GM_g \over rc^2} \right)  \left( dr^2 + r^2 d\theta ^2 + r^2 \sin^2 \theta d\varphi^2 \right) \ , 
\end{equation}
which nicely connects with the static pN metric explored later.
%and we also have
%\begin{equation}
%    \xi\approx r \left( 1 + {GM_g \over r c^2} %right)
%\end{equation}

Let us now work out the equations for hydrostatic equilibrium.  After a bit of algebra (not  provided here as the steps are standard) we arrive at an equation for the first metric potential
\begin{equation}
    {d \mu \over dr} = - 2  e^{-\mu/4} {G\bar M \over r^2 c^2} \ ,
   \label{dmu}
\end{equation}
where we have defined the mass $\bar M$ through
\begin{equation}
{d\bar M \over dr} = {4\pi \over c^2} \varepsilon e^{5\mu/4} r^2  \ .
\label{dMbar}
\end{equation}
The second potential follows from 
\begin{equation}
{d\nu\over dr}   =  2 e^{- (\nu+\mu)/2} {GM\over c^2 r^2}  \ , 
\label{dnu0}
 \end{equation}
where we have defined a different mass function, $M$, through
\begin{equation}
    {d M \over dr} =  4\pi  {\mathcal R} r^2  \ , 
    \label{dM}
\end{equation}
with
\begin{equation}
    {\mathcal R} = {1 \over c^2} \left(\varepsilon+3p\right) e^{(\nu+3\mu)/2} \ .
    \label{Rdef}
\end{equation}
Finally, the pressure follows from
 \begin{equation}
  {d  {\mathcal P}\over dr} =  - e^{-2(\nu+\mu)}{ G  M  {\mathcal R} \over r^2} \ ,
 \end{equation}
 where we have defined
 \begin{equation}
 {\mathcal P} = p e^{-\nu}  \ .
 \end{equation}

In these equations we have tried to use variables that make the results appear as ``neat'' as possible. This evidently includes two different masses, $\bar M$ and $M$. It is also  notable that, in this version of the equations neither of the metric potentials decouple from the pressure. This makes the isotropic problem---at least at first glance---rather different from the standard TOV equations. If we want to solve the equations as given above, then we need to provide values for both potentials at the centre of the star, $\mu(r=0)=\mu_0$ and $\nu_0$. In effect, the solution  involves a two-dimensional root-search, matching the integrated variables to the external metric (see  \cite{2015PhRvD..91j4030T} for relevant comments).  

At the surface of the star, $r=R$, we need both $\mu$ and its derivative to match continuously to the exterior (continuity of the first and second fundamental forms). We want to ensure that
\begin{equation}
    e^{\mu/4} = 1 + {GM_g\over 2Rc^2} 
    \label{mumatch}
\end{equation}
and 
\begin{equation}
     {d\over dr} \left( e^{\mu/4} \right) = - {GM_g\over 2R^2c^2} \ .
\end{equation}
Starting from the latter, for a chosen value of $\mu_0$, we arrive at the interior solution
\begin{equation}
    {d\over dr} \left( e^{\mu/4} \right) = -  {1\over 2} {G\bar M \over r^2 c^2}  \ .
\end{equation}
The matching at the surface then  leads to
\begin{equation}
    M_g = \bar M 
\end{equation}
and it follows from \eqref{mumatch} that
\begin{equation}
 e^{\mu(R)/4} = 1 + {G\bar M \over 2Rc^2}  \ .
\end{equation}
 This provides the condition needed to (iteratively) determine the value for $\mu_0$. 
However, there is a snag. In order to solve the equations we need the energy $\varepsilon(r)$, which follows from the solution for the pressure (and the assumed equation of state). Moreover, the location of the surface follows from $p(r=R)=0$. This then brings in the second, a priori unknown, central value $\nu_0$.

It would clearly be better if we could decouple at least one of the potentials from the equation for the pressure (as in the TOV case). One way to go about doing this is to define yet another mass function. With this in mind, we introduce 
 \begin{equation}
     \hat M = e^{-\nu/2} M \Longrightarrow {dp \over dr} = - (p+\varepsilon) e^{-\mu/2} {G\hat M \over r^2 c^2}
     \label{dphat}
 \end{equation}
 and 
 \begin{equation}
{d\hat M \over dr} 
 = {4\pi \over c^2} ( \varepsilon + 3p) e^{3\mu/2} r^2 -   e^{- \mu/2} {G\hat M^2 \over c^2 r^2}  \ .
 \label{dMhat}
 \end{equation}
 Working with $\hat M$ we decouple the second potential, $\nu$, from the pressure equation. This then allows us to solve the problem in two stages, where the first step only involves determining the appropriate value for $\mu_0$. Once this is done, the second potential follows from  \eqref{dnu0}--\eqref{Rdef}. Finally, the value of $\nu_0$ needs to be such that 
\begin{equation}
    e^{\nu(R)/2} = \left(  1 -  {GM_g \over 2Rc^2}  \right) e^{-\mu(R)/4} \Longrightarrow {GM_g \over 2 R c^2}  = 1 - e^{(2\nu(R) + \mu(R))/4} \ .
\end{equation}
Equivalently, this can be written
\begin{equation}
1 + e^{\nu/2} = 2 e^{-\mu/4} 
    \label{fudge}
\end{equation}
which  leads to
\begin{equation}
    \nu' e^{\nu/2} = -\mu' e^{-\mu/4} \ .
\end{equation}
Making use of the interior equations,  this condition corresponds to
\begin{equation}
      2 e^{- \mu/2} {GM\over c^2 R^2}    =  2  e^{-\mu/2} {G\bar M \over R^2 c^2} \Longrightarrow M=\bar M \ .
\end{equation}
 At the end of the day, a consistent solution requires
\begin{equation}
M(R)=\bar M(R) = M_g \ .
\end{equation}
The mass functions may differ inside the star, but they all lead to the same result at the surface---evidently in the spirit of the Newtonian virial relations.

\subsection{Scaling argument}

The arguments we have provided simplifies the strategy for solving for hydrostatic equilibrium in isotropic coordinates. Essentially, we need two root searches, first for $\mu_0$ and then for $\nu_0$. This is easier to implement than the original two-dimensional problem. However, the logic is still more involved than we are used to from the TOV problem. It is natural to ask if we can do better. It turns out that we can.

First, introduce
\begin{equation}
    \psi = e^{\mu/4}
\end{equation}
and rewrite \eqref{dmu} as
\begin{equation}
    {d \psi \over dr} = -  {G\bar M \over 2 r^2 c^2} \ .
\end{equation}
The other equations we need now take the form
\begin{equation}
{d\bar M \over dr} = {4\pi \over c^2} \varepsilon \psi^5 r^2  \ , 
\end{equation}
 \begin{equation}
    {dp \over dr} = - (p+\varepsilon) \psi^{-2} {G\hat M \over r^2 c^2}
 \end{equation}
 and 
 \begin{equation}
{d\hat M \over dr} 
 = {4\pi \over c^2} ( \varepsilon + 3p) \psi^6 r^2 -   \psi^{-2} {G\hat M^2 \over c^2 r^2} \ .
 \end{equation}
 
 Second, ask if the equations are invariant under some suitable rescaling. The argument proceeds as follows. Scale the variables in such a way that 
 \begin{equation}
     r \to \beta r\ , \quad \bar M \to \eta \bar M \ , \quad \psi \to \alpha \psi \ ,
 \end{equation}
 where $\beta$, $\alpha$ and $\eta$ are all constant. From the first two equations, we then see that we need to have
\begin{equation}
    {\alpha \over \beta} = {\eta \over \beta^2} \quad  \mbox{and} \quad {\eta \over \beta} = \alpha^5 \beta^2 \ ,
\end{equation}
leading to 
\begin{equation}
    \beta = {1\over \alpha^2} \quad \mbox{and} \quad \eta = {1\over \alpha} \ .
\end{equation}
Next, in order for the pressure equation to be invariant,  we must have
\begin{equation}
    \hat M \to \beta \alpha^2 \hat M = \hat M \ , 
\end{equation}
which is consistent with the final equation, as well. 

In essence, once we have a solution to the equations we can find other solutions by scaling. We can use this ``trick'' to ensure the matching at the star's surface. Starting from an arbitrary value for $\mu_0$, integrate the equations up to the point where $p(R_*)=0$ and  $\bar M(R_*) = \bar M_*$. We also obtain the potential $\psi_*$.

At the surface of the star the true values of the variables should be such that
\begin{equation}
    \psi(R) =  1 + {G\bar M\over 2Rc^2} \ .
\end{equation}
Scaling the numerical solution, we require
\begin{equation}
\alpha \psi_* = 1 + {G \eta \bar M_* \over 2 \beta R_* c^2} = 1 + {G  \alpha \bar M_* \over 2 R_* c^2}  
\end{equation}
so we must have
\begin{equation}
    \alpha = \left( \psi_*- {G  \bar M_* \over 2 R_* c^2} \right)^{-1} \ .
\end{equation}
This provides the correct central value for $\psi$ (and hence $\mu$). The other variables for the stellar model follow from the appropriate scaling.  

A similar argument applies to the second potential. Introduce
\begin{equation}
     N = e^{\nu/2}  
\end{equation}
and rewrite \eqref{dnu0} as 
\begin{equation}
{d N \over dr}   =  \psi^{-2} {GM\over c^2 r^2}  
\label{dN}
 \end{equation}
along with
\begin{equation}
    {d M \over dr} =  {4\pi r^2 \over c^2} \left(\varepsilon+3p\right)  N \psi^6 \ .
\end{equation}
Assuming a scaling such that 
\begin{equation}
     N \to \zeta  N
\end{equation}
we see that we must have 
\begin{equation}
    M \to \beta \left( \beta^2 \zeta \alpha^6 \right) M = \zeta M \ ,
\end{equation}
which also leaves the equation for $ N$ invariant. The surface matching, adding a suitable central value for $\nu_0$ to the previously calculated model, now requires
\begin{equation}
1 +  N = {2 \over \psi}
\end{equation}
and the scaled solution leads to 
\begin{equation}
    \zeta  = {1\over  N_*} \left( {2\over \alpha \psi_*} - 1 \right) = {1\over  N_*} \left[ {2\over \psi_*} \left( \psi_*- {G  \bar M_* \over 2 R_* c^2} \right) - 1 \right] = {1\over  N_*} \left( 1 - {G  \bar M_* \over \psi_*  R_* c^2}\right) \ ,
\end{equation}
which fixes the value of $\zeta$ and completes the stellar model. Alternatively, we can use the fact that the final result should be consistent with $M(R)=\bar M(R)$, which leads to 
\begin{equation}
   \zeta = \alpha {\bar M_* \over M_*} =  \left( \psi_*- {G  \bar M_* \over 2 R_* c^2} \right)^{-1} {\bar M_* \over M_*} \ .
\end{equation}

 \subsection{Truncated isotropic models}
 
 The scaling argument leads to an elegant prescription for building stars using isotropic coordinates, notably not involving a root search at all. It also helps us understand what happens if we expand the equations to some power of $1/c^2$, as in the case of the truncated TOV equations from section~\ref{tovtrunc}. This turns out to be an instructive exercise---bringing out issues relevant also for the pN discussion to follow. 
 
 Focussing on the mass-radius relation, we first 
 note that we can leave out the step that determines the second metric potential $\mathcal N = e^{\nu/2}$. That is, we focus on the system of equations \eqref{dmu}--\eqref{dMbar} and \eqref{dphat}--\eqref{dMhat}. We do, however, need to commit to an expansion for the potential $\psi = e^{\mu/4}$. Intuitively, as it connects with the pN logic, it is natural to assume that
\begin{equation}
e^\mu \equiv e^{2\bar U/c^2} \approx 1 + {2 \bar U \over c^2} \Longrightarrow {d\mu \over dr} \approx e^{-\mu} {2 \over c^2} {d\bar U \over dr}  \ , 
\end{equation}
where $\bar U$ plays the role of the gravitational potential.

In order to check to what extent additional assumptions make a difference, we will compare two slightly different models. First, 
suppose we do not separate the energy into mass and internal energy parts, but leave $\varepsilon$ and the combination with the pressure as they are. Then we need to keep in mind that $\varepsilon/c^2 = \mathcal O(1)$ to get
\begin{equation}
{d\bar M \over dr} \approx {4\pi \over c^2} \varepsilon r^2 \left( 1 + {5\bar U \over 2 c^2} \right)  \ , 
\label{dMbarnew}
\end{equation}
\begin{equation}
    {d \mu \over dr} = - 2  e^{-\mu/4} {G\bar M \over r^2 c^2} 
    \Longrightarrow 
   {d\bar U \over dr} \approx  - e^{3\mu/4} {G\bar M \over r^2 } \approx - {G\bar M \over r^2} \left( 1 + {3\bar U \over 2 c^2} \right) \ ,
\end{equation}
\begin{equation}
 {d\hat M \over dr} 
 \approx {4\pi \over c^2} ( \varepsilon + 3p)  r^2 \left( 1+ {3\bar U \over c^2} \right)  -  {G\hat M^2 \over c^2 r^2} \ , 
 \end{equation}
and
\begin{equation}
    {dp \over dr} \approx - (p+\varepsilon)  {G\hat M \over r^2 c^2} \left( 1- {\bar U \over c^2} \right) \ .
    \label{dpbarnew}
 \end{equation}
Once these equations are integrated to the surface $p=p(R)=0$, we need to match to the external potential at the surface. We then need
\begin{equation}
    e^{\mu/4} = 1 + {GM_g\over 2Rc^2}  
\Longrightarrow 
\bar U(R) = {GM_g\over R} \ ,
\end{equation}
where we have to keep in mind that we do not yet know $M_g$,
but 
\begin{equation}
     {d\over dr} \left( e^{\mu/4} \right) = - {GM_g\over 2R^2c^2} \Longrightarrow \bar M(R) = M_g \ .
\end{equation}
It follows that the central value of the potential $(\bar U_0$) is  fixed by the condition
\begin{equation}
    \bar U(R) = {G \bar M(R) \over R} \ .
    \label{surfcon}
\end{equation}

At this point---and for future reference---it is important to note that the expansion breaks the scaling argument from the previous section. Basically, we now have the freedom to change the central value of the potential $\bar U$ by adding a constant. The upshot of this is that the correct central value $\bar U_0$
 must be iteratively determined by a root search.
 
In the second model, we expand the energy in the usual way. Using \eqref{toten} we have
\begin{equation}
{d\bar M \over dr} = 4\pi \rho r^2 \left( 1 + {\Pi \over c^2} \right)  \left( 1 + {5\bar U \over 2 c^2} \right) \approx  4\pi \rho r^2 \left[ 1 + {1\over c^2} \left(\Pi   + {5\bar U \over 2 } \right) \right] \ , 
\label{dMbar2}
\end{equation}
\begin{equation}
    {d \mu \over dr} = - 2  e^{-\mu/4} {G\bar M \over r^2 c^2}
    \Longrightarrow 
   {d\bar U \over dr} \approx  - e^{3\mu/4} {G\bar M \over r^2 } \approx - {G\bar M \over r^2} \left( 1 + {3\bar U \over 2 c^2} \right) \ ,
\end{equation}
\begin{equation}
 {d\hat M \over dr} 
 = {4\pi \over c^2} ( \varepsilon + 3p) e^{3\mu/2} r^2 -   e^{- \mu/2} {G\hat M^2 \over c^2 r^2} \\
 4\pi \rho r^2  \left[ 1 + {1\over c^2} \left(\Pi + {3p\over \rho} + 3\bar U \right) \right] -  {G\hat M^2 \over c^2 r^2} 
 \end{equation}
and
\begin{equation}
    {dp \over dr} = - (p+\varepsilon) e^{-\mu/2} {G\hat M \over r^2 c^2} \approx - {G\hat M \rho \over r^2} \left[ 1 + {1\over c^2} \left(\Pi + {p\over \rho} - \bar U  \right) \right] \ .
    \label{dpbar2}
 \end{equation}
 The matching condition at the surface remains as before.
 
\begin{figure}
    \centering
    \includegraphics[width=0.75\textwidth]{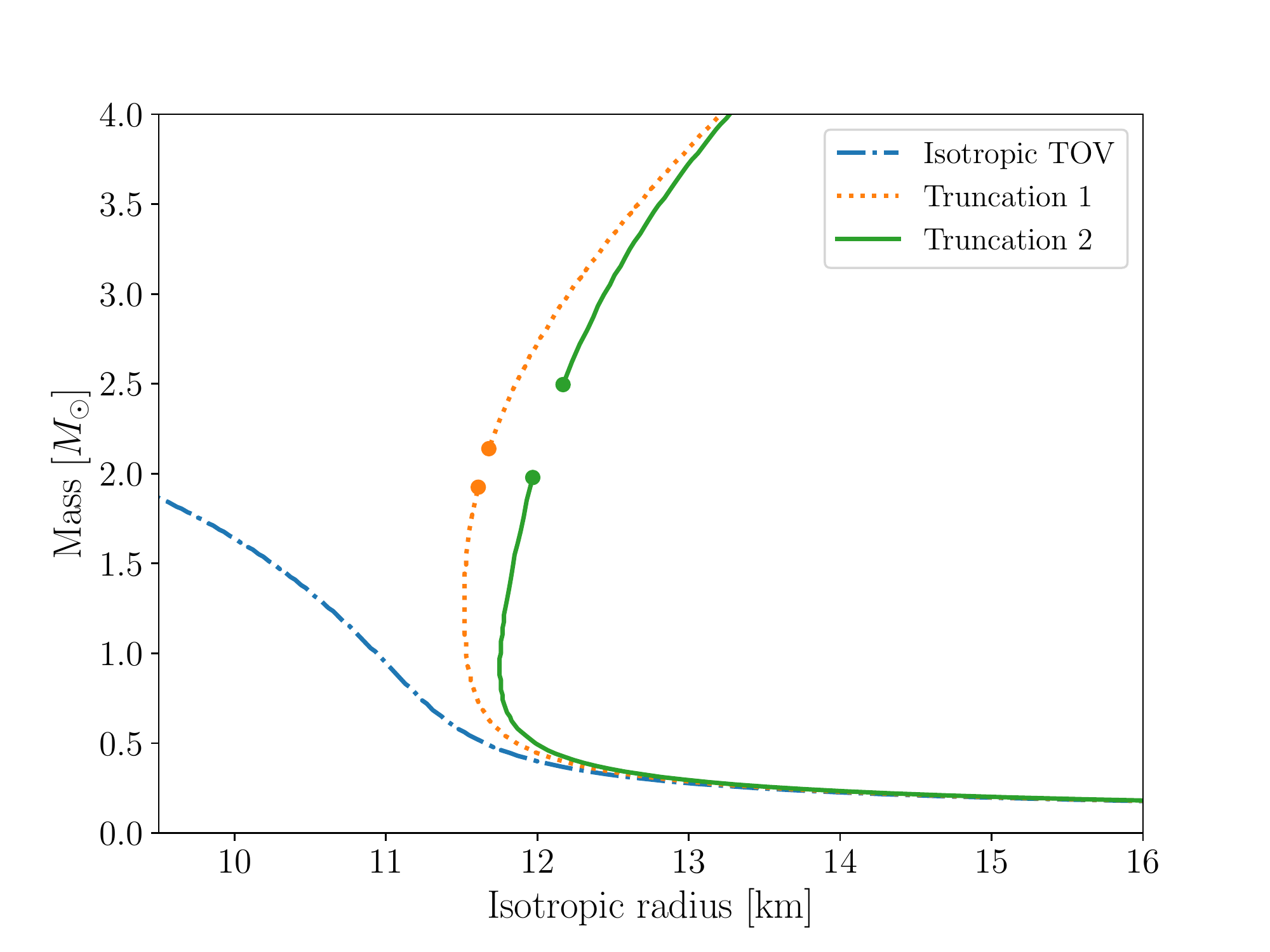}
    \caption{Mass-radius curves for the two first order truncations of the equations for hydrostatic equilibrium in isotropic coordinates, \eqref{dMbarnew}--\eqref{dpbarnew} (orange dotted) and \eqref{dMbar2}--\eqref{dpbar2} (green solid). For reference we also so the results from the isotropic TOV equations (blue dot-dash). Data correspond to the BSk24 equation of state.}
    \label{trunc1}
\end{figure}

\begin{figure}
    \centering
    \includegraphics[width=0.75\textwidth]{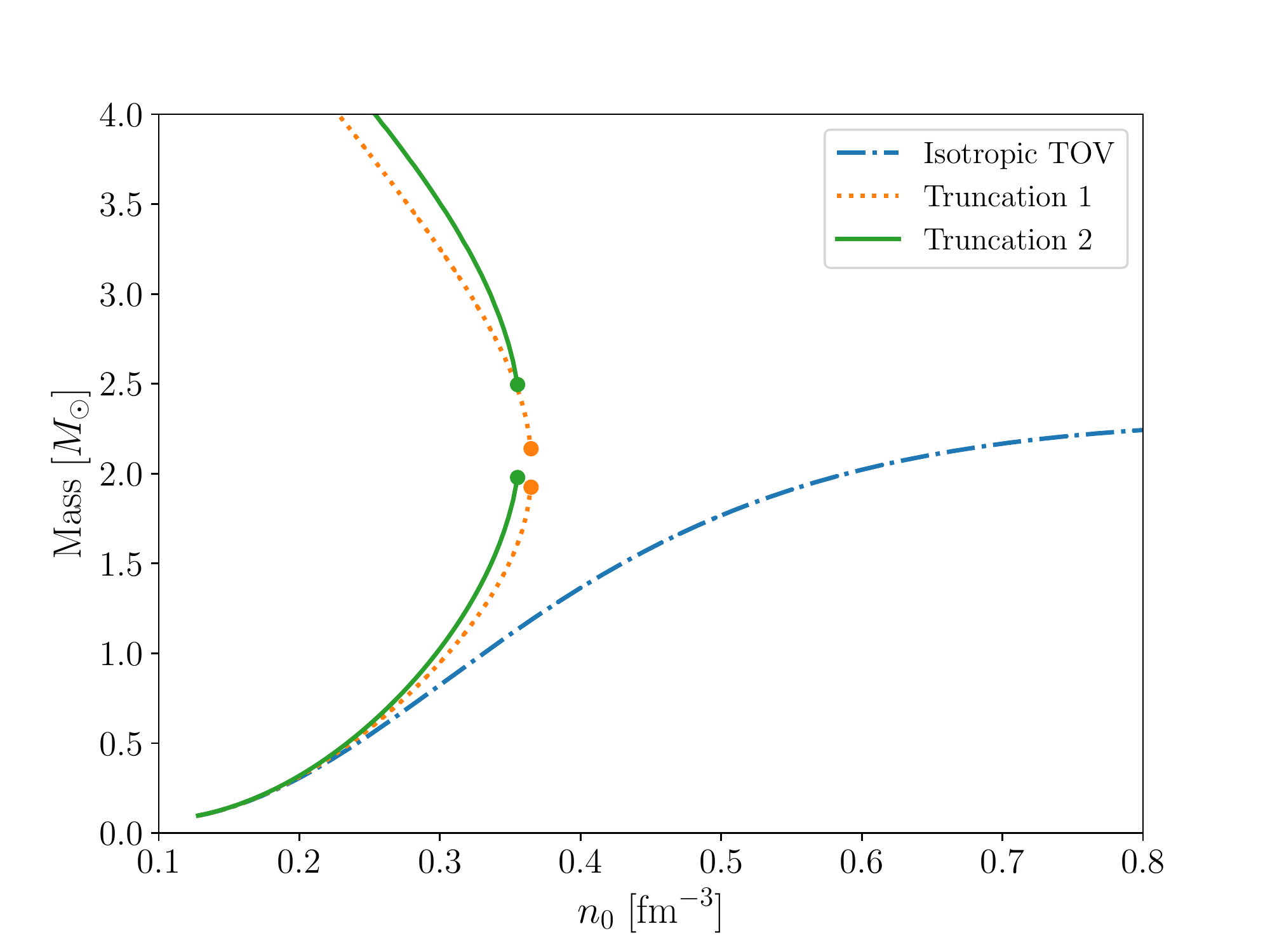}
    \caption{ Mass against central number density ($n_0$) curves for the two first order truncations of the equations for hydrostatic equilibrium in isotropic coordinates, \eqref{dMbarnew}--\eqref{dpbarnew} (orange dotted) and \eqref{dMbar2}--\eqref{dpbar2} (green solid). For reference we also so the results from the isotropic TOV equations (blue dot-dash). Data correspond to the BSk24 equation of state.}
    \label{trunc2}
\end{figure}

There should be nothing particularly controversial here. The steps we have taken seem natural. However, when we solve either set of truncated equations---\eqref{dMbarnew}--\eqref{dpbarnew} or \eqref{dMbar2}--\eqref{dpbar2}--- the result is surprising. Basically, we find that there is a maximum density beyond which no stellar model can be found. Typical results (again for the BSk24 equation of state) are provided in figures~\ref{trunc1} and \ref{trunc2}. The behaviour is very different from the solution to the full isotropic version of the TOV equations (which leads to a mass-radius curve identical to the original TOV solution from figure~\ref{TOVfig} once we translate the coordinate value of the radius). What is happening here is that the expansion in $1/c^2$ has made the eigenvalue problem for $\bar U_0$ nonlinear. At low densities there are two roots. One root brings us back to the Newtonian limit and the other leads to very massive stars with a low value of the central density. As the density increases, these two sequences approach each other until---at a specific maximum value of the density---they merge. Beyond this density there are no solutions. In figures~\ref{trunc1} and \ref{trunc2}, we have ended our calculation when the two roots are closer than a set numerical resolution. This leaves a gap between the two sequences. While we could make this gap smaller by pushing the calculation to higher precision, we have left  it like this to illustrate the main take-home message. 

On the one hand, one may argue that the result is what it is. We should not be surprised that the truncated solution differs from the full ``nonlinear'' one when  higher order terms in the $1/c^2$ expansion play a prominent role. The result is just a warning that we should not use the truncated models in the regime where they do not apply. The conclusions are nevertheless disturbing, especially as we are about to turn to the pN problem, which is similar in spirit. There are  many situations where one might want to use an approximate neutron-star model, even if it may not be particularly precise. What we are learning is that such approximations may have less attractive features. Consider the simple fact that, for a given equation of state we can build neutron-star models up to the maximum mass (and beyond) using the TOV equations (in Schwarzschild or isotropic coordinates, whichever may be preferred). Similarly, we can find solutions to the Newtonian equations for any given central density. These solutions may not provide a very accurate representation of neutron stars (there is no maximum mass, etcetera), but they exist. The ``back-bending'' of the  curves in figures~\ref{trunc1} and \ref{trunc2} beyond some density seems an unintuitive and not particularly welcome feature of the truncated problem. Having said that, if all we want is a reasonably good approximation then it is worth noting that we can use the truncated models to build canonical neutron stars with a mass of $1.4M_\odot$ with errors in the radius less than 1~km. This level of error is notably similar to the current observational constraints, obtained from multimessenger data (including the NICER radius results for the low-mass pulsar J0030+0451 and the high-mass system J0740+6620) \cite{2021ApJ...918L..29R}. This suggest that, while the pN models may be viewed as ``good enough'' at the moment, we have to do better in the future.

\section{Formulating the pN problem}
\label{sec:formulating}

At this point, we could continue to explore the
weak-field limit of the isotropic TOV equations with the aim to gain further insight into the $1/c^2$ expansion. Rather than doing this, we will move straight to the pN version of the problem. The main difference is that, instead of truncating the relativistic equations we will build on the pN formulation of hydrodynamics. 
In doing this we heavily rely on the exhaustive discussion in the recent monograph by \citet{PWbook}. In this description the baryon number density plays central role, so it makes sense to begin by explaining how this comes about. 

\subsection{Baryon number conservation}

Suppose we start from the perfect fluid stress-energy tensor
\begin{equation}
    T^{ab} = {1\over c^2} ( \varepsilon+ p)  u^a u^b + p g^{ab} \ , 
\end{equation}
where $u^a$ is the fluid four velocity, along with the baryon number conservation law
\begin{equation}
    \nabla_a (n u^a) = 0 \Longrightarrow \frac{1}{\sqrt{-g}} \partial_a \left( \sqrt{-g} nu^a \right) = 0\ .
\end{equation}
Now introduce a different observer such that
\begin{equation}
u^0 = \gamma c\ , \quad u^i = \gamma v^i  \ , 
\end{equation}
where $\gamma = u^0/c $ follows from the normalisation condition
\begin{equation}
    g_{ab} u^a u^b = - c^2  \ .
\end{equation}

We then have (keeping in mind that $x^0= ct$)
\begin{equation}
{1\over c} \partial_t \left(m_B \sqrt{-g} n u^0 \right) + \partial_i \left( m_B \sqrt{-g} n u^i \right) = 0  
\end{equation}
or 
\begin{equation}
    \partial_t \rho^* + \partial_i (\rho^* v^i) = 0  \ , 
    \label{conlaw}
\end{equation}
where $\rho = m_B n$ is the baryon mass density (as before) and we have defined 
\begin{equation}
    \rho^* = \sqrt{-g} \rho {u^0 \over c} =\sqrt{-g} \rho \gamma  \ .
\end{equation}

So far, the description is fully relativistic so equation \eqref{conlaw} should hold at all pN orders \cite{PWbook}. This makes $\rho^*$ an attractive variable to work with.

\subsection{The pN formulation}

Now consider a formal expansion in  inverse powers of $c$. We then need the pN form for the metric (there are different choices, depending on the chosen gauge \cite{1993tegp.book.....W}). Here we take equations~(8.2) from \cite{PWbook}, which are expressed in the harmonic gauge,  as our starting point. We then have 
\begin{equation}
    g_{00} = - 1 + {2U\over c^2} + {2\over c^4} \left( \Psi - U^2 \right)  + \mathcal O(c^{-5}) \ ,
\end{equation}
\begin{equation}
    g_{0j} = - {4U_j \over c^3}  + \mathcal O(c^{-5}) 
\end{equation}
and
\begin{equation}
    g_{jk} = \delta_{jk} \left( 1 + {2U \over c^2} \right) +  \mathcal O(c^{-4}) \ ,
    \label{gij}
\end{equation}
where we have introduced the potentials\footnote{Note that the use of $\rho^*$ here means that we are---strictly speaking---defining the potentials in such a way that they contain pN terms. The argument for this becomes clear(er) when you consider the different mass functions we introduce later. }
\begin{equation}
    \nabla^2 U = - 4\pi G \rho^*
\end{equation}
and
\begin{equation}
\nabla^2 U_i = -4\pi G \rho^* v_i \ , 
\end{equation}
where $v_i$ is the fluid (three-)velocity.
We also have 
\begin{equation}
    \Psi = \psi + {1\over 2} \partial_{tt} X 
\end{equation}
where
\begin{equation}
   \nabla^2  \psi= -4\pi G \rho^* \left({3\over 2} v^2 - U + \Pi + {3p\over \rho} \right) \ , 
\end{equation}
and the superpotential $X$ is determined by
\begin{equation}
 \nabla^2 X = 2U \Longrightarrow   X = G \int \rho^* |\vec x - \vec x'| d^3x'  \ .
\end{equation}

Given these definitions we find that 
\begin{equation}
    \gamma = \left( 1 - {v^2 \over c^2} - {2U\over c^2}  \right)^{-1/2} \approx 1 + {v^2 \over 2c^2} + {U\over c^2 } +  \mathcal O(c^{-4}) \ .
\end{equation}

If we also note that
\begin{equation}
    \sqrt{-g} = 1 + {2U\over c^2} + \mathcal O(c^{-4}) \ ,
\end{equation}
we see that
\begin{equation}
    \rho^* \approx \left( 1 + {2U\over c^2}\right) \left( 1 + {v^2 \over 2c^2} + {U\over c^2 } \right)\rho\approx \left( 1 + {v^2 \over 2c^2} + {3U\over c^2 } \right)\rho 
\end{equation}
and, for future reference, it is worth noting that
\begin{equation}
    \sqrt{-g} \approx \left(1 - {U\over c^2} \right) \sqrt{^{(3)}g} \approx  \left(1 - {U\over c^2} \right) \left(1 + {3U\over c^2} \right) =  1 + {2U\over c^2} \ .
\end{equation}

\subsection{Static fluid configurations}

For a static fluid configuration, we ignore all time derivatives and also set $v^i=0$. Noting that---this is evident from \eqref{gij}---the pN metric is expressed in isotropic coordinates,  letting the corresponding radial coordinate be $r$ (as before) and following the steps in \cite{PWbook} we  arrive at the equation for hydrostatic balance at 1pN order:
\begin{equation}
    \left[ 1 -  {1\over c^2} \left( \Pi + U  + {p \over \rho^* }  \right)\right]{dp\over dr} = \rho^* \left( 1 - {1\over c^2} 4U \right) {dU \over dr} +{\rho^*\over c^2}  {d\psi\over dr} \ .
    \label{preseq1}
\end{equation}
That is, we have
\begin{equation}
    {dp \over dr} 
= \rho^* \left[ 1 +   {1\over c^2} \left( \Pi - 3U  + {p \over \rho^* }  \right)\right] {dU \over dr} +{1\over c^2} \rho^* {d \psi \over dr} \ ,
\label{equil1}\end{equation}
where
\begin{equation}
    \rho^* = \left( 1 + {3U\over c^2 } \right)\rho \ .
\end{equation}
We also note that $U_i=0$ and $\Psi=\psi$ so we do not have to worry about the superpotential in the following.

Conveniently, we may write the equations we need to solve as a coupled set of first order differential equations. First we have
\begin{equation}
    {dM_B \over dr} = 4\pi \rho^* r^2  \ , 
\end{equation}
with $M_B$ representing the baryon mass (recall $\rho=m_B n$). This is evident since
\begin{equation}
    M_B = 4\pi \int  \rho^* r^2 dr = \int \rho \sqrt{^{(3)}g}\, d^3x = m_B \int n dV \ , 
\end{equation}
where $dV$ is the (three-)volume element for the pN metric.
Next it follows that  
\begin{equation}
 {1\over r^2} {d\over dr} \left( r^2 {dU \over dr} \right) = - 4\pi G \rho^* \Longrightarrow {dU \over dr} = - {GM_B\over r^2}  \ .
\end{equation}

In the vacuum outside the star, we must have---regardless of which mass we use to source the gravitational potential (see later):
\begin{equation}
{d\over dr} \left( r^2 {dU \over dr} \right)  = 0  \Longrightarrow  {dU \over dr} = - {A\over r^2} \Longrightarrow U = {A\over r} \ .
\end{equation} 
The constant $A$ depends on the source of the potential, but the surface matching condition does not. We only need to make sure that 
\begin{equation}
    A= -r^2 {dU\over dr} = rU \quad \mbox{at}\ r=R \ .
\end{equation}
That is, we need
\begin{equation}
    r{dU\over dr} + U = 0 \quad \mbox{at} \ r=R \ ,
\end{equation}
to hold for all pN models.

We also have the two equations
\begin{equation}
    {d\psi \over dr} = - {G \mathcal N \over r^2} 
\end{equation}
and
\begin{equation}
   {d \mathcal N \over dr} = 4\pi \rho^* r^2  \left(  \Pi - U  + {3p\over \rho^*} \right) \ .
\end{equation}
With these definitions \eqref{equil1} takes the form
\begin{equation}
       {dp \over dr} 
= - {G \rho^* \over r^2} \left\{ M_B +   {1\over c^2} \left[ \left( \Pi - 3U  + {p \over \rho^* }  \right)M_B + \mathcal N \right] \right\} \ .
\label{PWpressure}
\end{equation}
These are the pN equations as given in exercise 8.5 of \cite{PWbook}. There are no great revelations here, but three comments are in order. First,  the equations we have written down clearly include higher order pN terms. This is not a ``problem'' as long as we are truly in the pN regime, where higher order terms in $1/c^2$ are small, but we have already seen from the example of the truncated TOV equations that such higher order terms may significantly impact on the  neutron-star models. Inevitably, the inclusion of higher order terms renders the modelling more ad hoc than one might be comfortable with. Second, as we need to solve for the potential $U$ we need to pay attention to the fact that the central value of the potential is not known ab initio (just as in the isotropic case). It needs to be solved for by matching at the surface. The third point relates to the equation of state. 

In equilibrium, neutron-star matter can be described in terms of a single parameter, usually taken to be the baryon number density. Thermodynamics then provides the Gibbs relation
\begin{equation}
    p + \varepsilon = n {d\varepsilon \over dn} \ , 
\end{equation}
or
\begin{equation}
   {d\varepsilon \over dn} = {p+\varepsilon \over n} \Longrightarrow   {d\varepsilon \over d\rho} = {p+\varepsilon \over \rho}  \ .
   \label{gibbs}
\end{equation}
In terms of the specific internal energy per unit mass $\Pi$, it follows that 
\begin{equation}
    d\Pi - {p \over \rho^2 } d\rho = 0 \ .
    \label{1stlaw}
\end{equation}
This leads us to issue a word of caution regarding the variable $\rho^*$. It may seem tempting, given the form of the pN equations we have written down to think of the matter equation of state as a relation $p=p(\rho^*)$. Mathematically, this might make sense but physically it is problematic. The matter description is developed in a local inertial frame and therefore the equation of state cannot be cognizant of the spacetime geometry. It should not depend on the potential $U$. 

\begin{figure}
    \centering
    \includegraphics[width=0.75\textwidth]{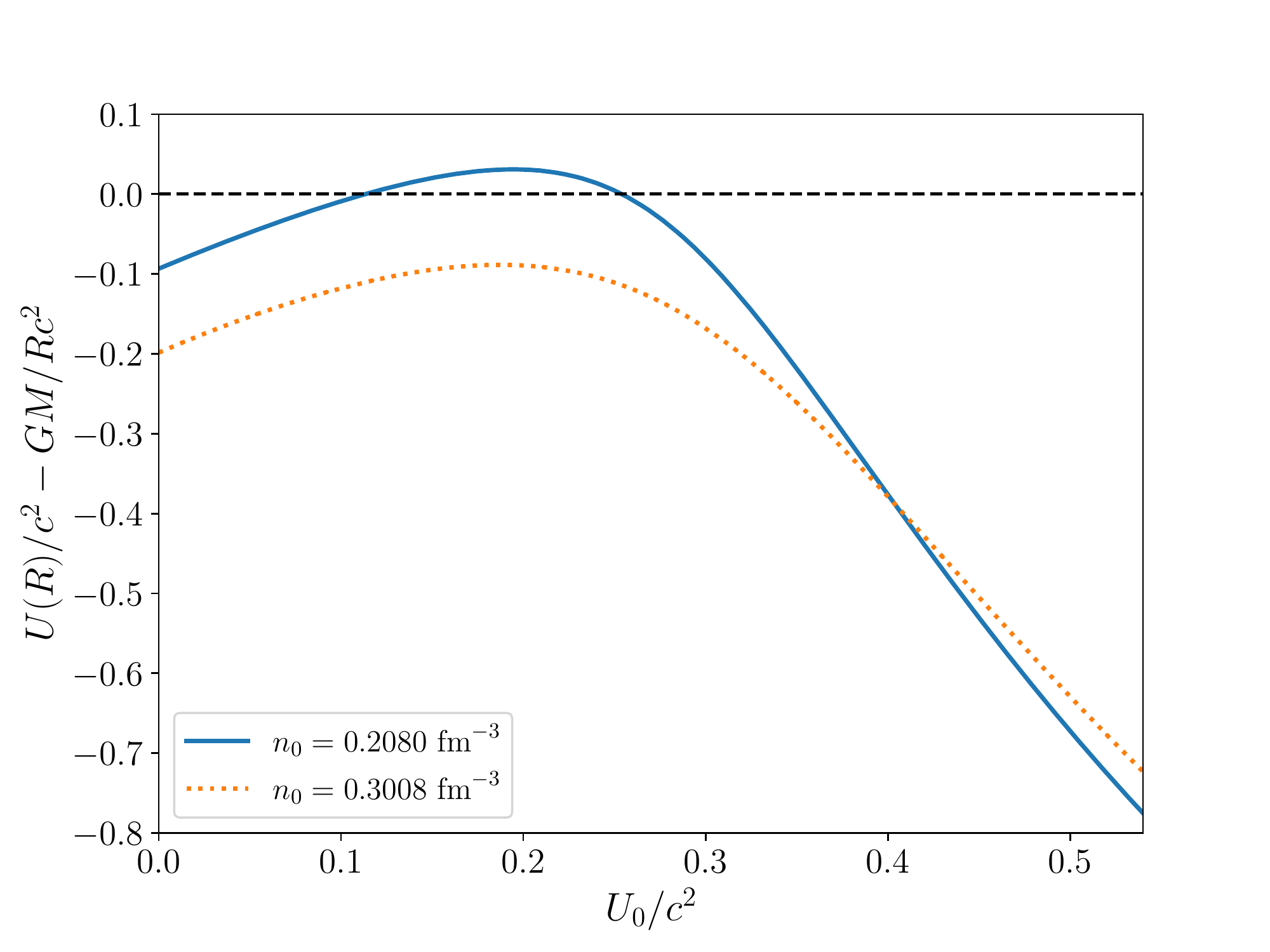}
    \caption{Demonstration that there is a maximum central density ($n_0$) beyond which  the internal gravitational potential cannot be matched to the exterior in the pN  scheme from \cite{PWbook}. Below a certain density the problem for the central value $U_0$ has two roots (the solid blue curve crosses zero at two points), but above a critical density there are no roots (orange dotted curve). This accords with the results for the isotropic TOV equations from figure~\ref{trunc2}. Data correspond to the BSk24 equation of state.}
    \label{nosoln}
\end{figure}

Let us now solve the pN equations in this form. When we do this, we find---perhaps not surprisingly---that the solution has the same behaviour as in the case of the truncated isotropic equations. Beyond some maximum central density there are no solutions. This is illustrated by the results in figure~\ref{nosoln} which shows the surface matching condition as function of the central value of the gravitational potential. At lower densities the problem has two roots (and hence there will be two stars with the same central density), but at higher densities there are no solutions. 

\begin{figure}
    \centering
    \includegraphics[width=0.75\textwidth]{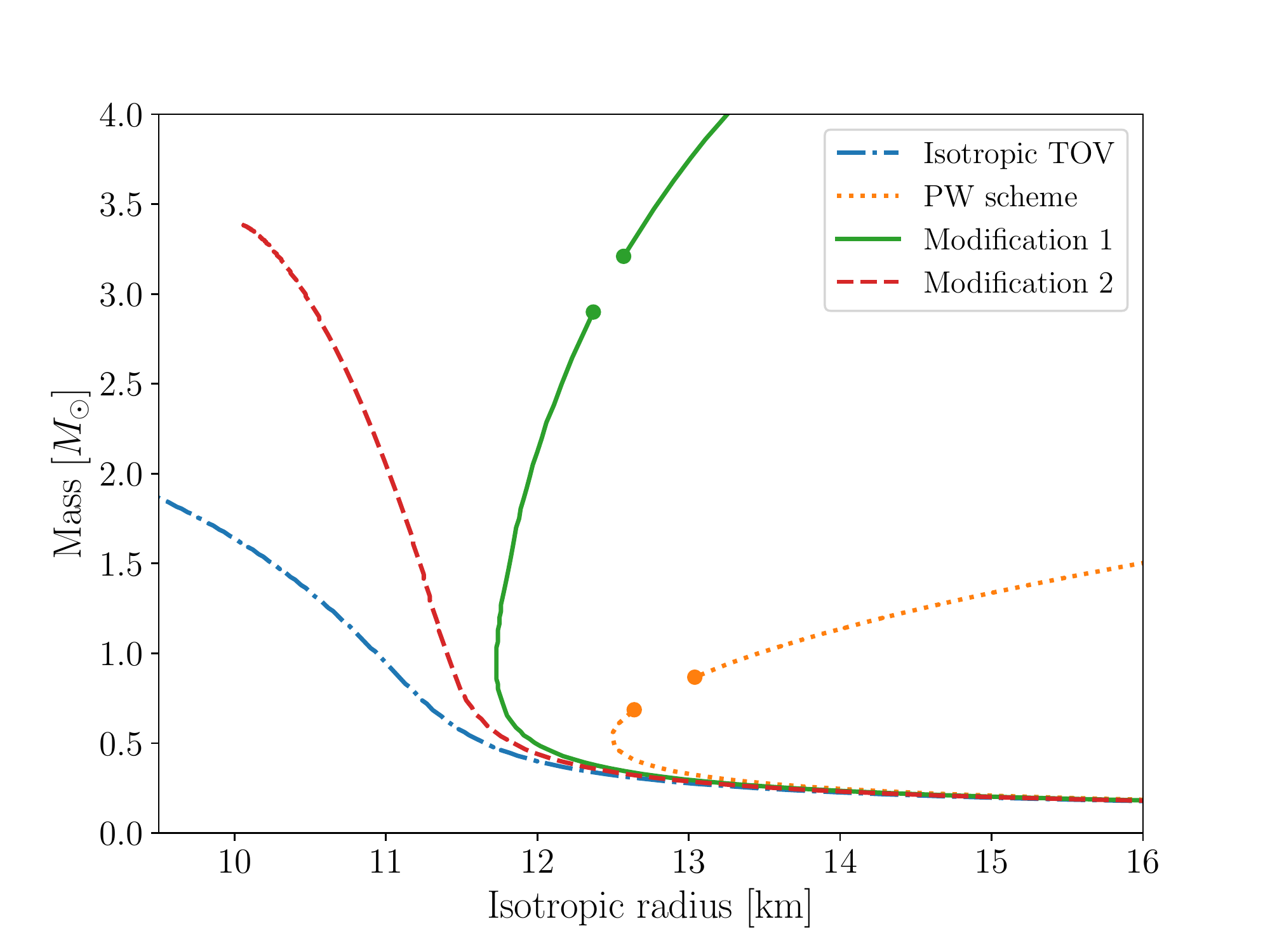}
    \caption{Mass-radius curves for the pN schemes \eqref{PWpressure} (orange dotted) and \eqref{truepN} (green solid). Also shown are results obtained from \eqref{shibeq} (red dashed). For comparions we also show the results from the isotropic TOV equations (blue dot-dash). Data correspond to the BSk24 equation of state.}
    \label{PWfig}
\end{figure}

\begin{figure}
    \centering
    \includegraphics[width=0.75\textwidth]{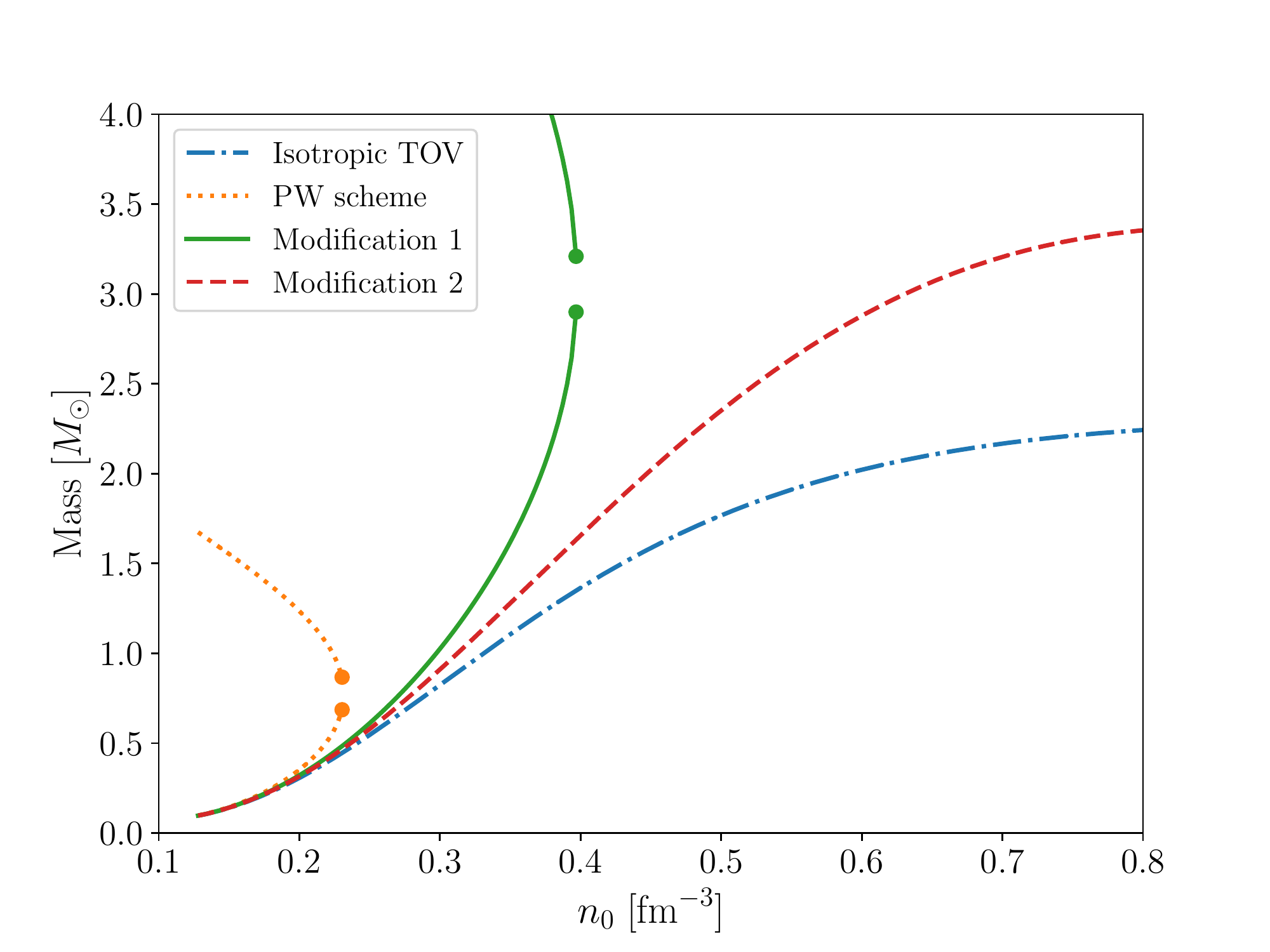}
    \caption{Mass against central number density ($n_0$) curves for the pN schemes \eqref{PWpressure} (orange dotted) and \eqref{truepN} (green solid). Also shown are results obtained from \eqref{shibeq} (red dashed). For comparions we also show the results from the isotropic TOV equations (blue dot-dash). Data correspond to the BSk24 equation of state.}
    \label{PWfig2}
\end{figure}

In order to fix the problem, we can try to remove some of the higher order pN terms. For example, it is easy to see that we may  rewrite \eqref{PWpressure} (equivalent at 1pN order) as 
\begin{equation}
       {dp \over dr} 
= - {G \rho \over r^2} \left\{ M_B +   {1\over c^2} \left[ \left( \Pi  + {p \over \rho }  \right)M_B + \mathcal N \right] \right\} \ .
\label{truepN}
\end{equation}
 Results from this equation are slightly better, see figure~\ref{PWfig}, but the problem of the two roots remains. We also show how the mass depends on the baryon number density in figure~\ref{PWfig2}. Noting that the radius of a $1.4M_\odot$ star is now closer to the correct result, it is evident that the inclusion (or not) of higher order pN terms can make a substantial difference. 

The two examples we have provided inevitably prompts us to ask if there is an optimal pN model. This is a tricky issue, given the somewhat arbitrary choices involved. Still, it is relevant to consider another couple of options.

An alternative would be to take the lead from \citet{1997PhRvD..55.6019S},  go back to \eqref{preseq1} and rewrite it as (again, consistently to order $1/c^2$)
\begin{equation}
      \left[ 1 -  {1\over c^2} \left( \Pi  + {p \over \rho }  \right)\right]{dp\over dr} = - {G \rho \over r^2} \left( M_B +{1\over c^2} \mathcal N  \right) \ .
\end{equation}
Instead of expanding this equation, we then divide\footnote{This is obviously not true to the spirit of the pN expansion, but one may argue that we  abandon such purity as soon as we break the strict ordering of the expansion anyway.} by the prefactor on the left-hand side. This then leads to
\begin{equation}
       {dp \over dr} 
= - {G \rho \over r^2} \left[ 1 -    {1\over c^2} \left( \Pi   + {p \over \rho }  \right) \right]^{-1} \left( M_B + {1\over c^2}  \mathcal N \right) \ .
\label{shibeq}
\end{equation}
Finally, expanding this result in  pN orders brings us back to \eqref{truepN}. Results obtained from \eqref{shibeq} are also shown in figure~\ref{PWfig}. This model is notably better. In fact, as can be seen from the results for polytropes in \cite{1997PhRvD..55.6019S} the model also leads to a maximum-mass configuration. This is an attractive feature, even though the predicted maximum mass is much too high. In principle, the results obtained from \eqref{shibeq} are attractive but one may have to view this with some caution as it is not clear that a strategy that involves dividing by terms involving pN corrections is ``helpful'' once we turn to dynamical problems.

\begin{figure}
    \centering
    \includegraphics[width=0.75\textwidth]{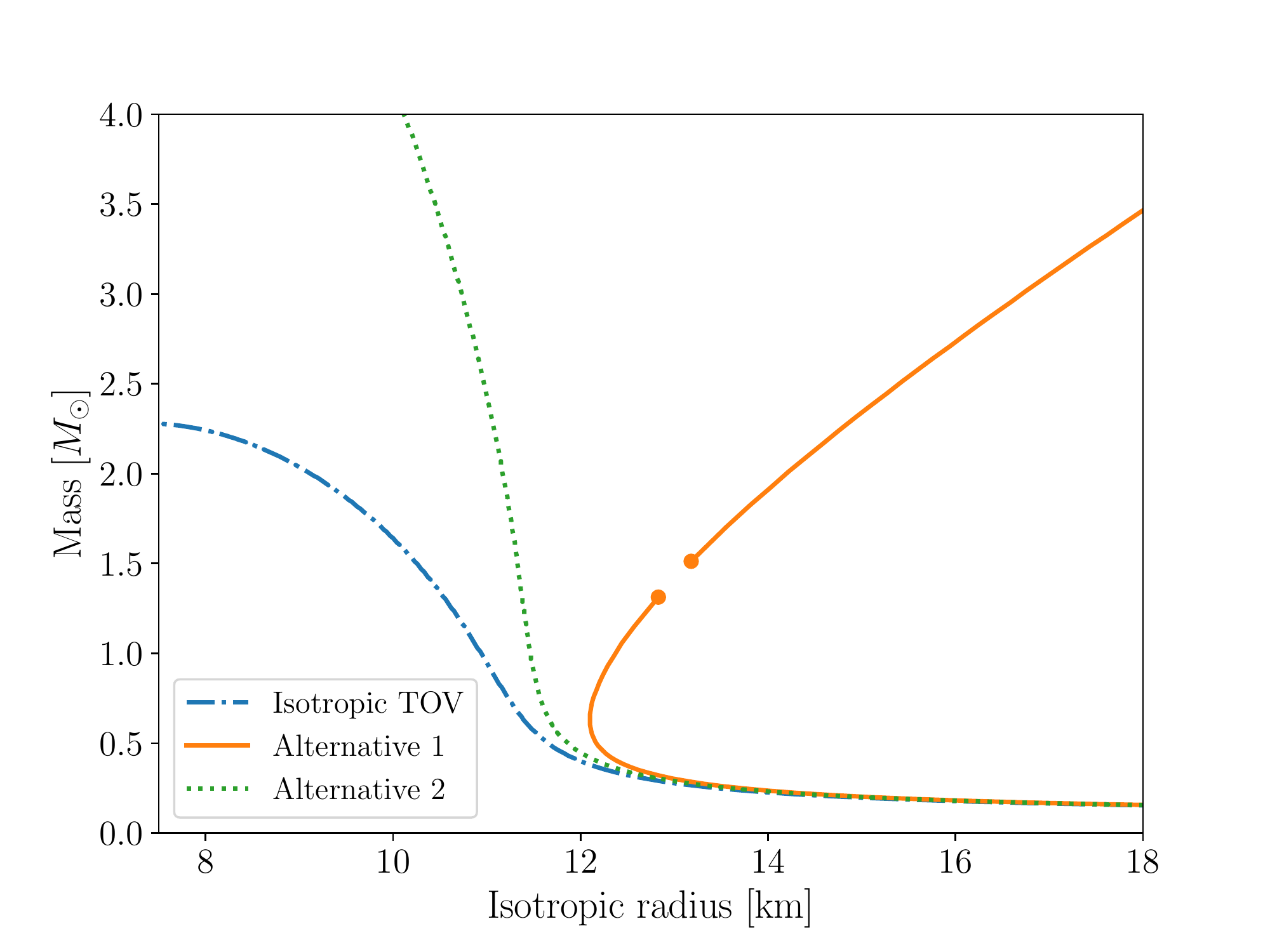}
    \caption{Comparing mass-radius curves obtained for alternative pN models from \eqref{BDSone} (orange solid) and \eqref{BDStwo} (green dotted) to the original TOV solution (blue dot-dash). Data correspond to the BSk24 equation of state.}
    \label{pwadjust}
\end{figure}

\subsection{Different mass functions}

A useful lesson---apparent already from the discussion of the truncated TOV equations in section~\ref{tovtrunc}, see also \cite{2015PhRvD..92b4056G}---is that we may use different mass functions to our advantage. Our goal is to explore, given the freedom afforded by the pN approximation, how different definitions at 1pN enable us to construct more, or (indeed) less, accurate neutron-star models. With this in mind, we note that the pN variables we have used may not be the most ``intuitive''. For example, comparing to the relativistic result one would not expect the gravitational potential to be sourced by the baryon mass. Given this, let us introduce
\begin{equation}
   M = M_B + {1\over c^2 }\mathcal N 
\end{equation}
so that
\begin{equation}
    {d M \over dr} = 4\pi \rho^* \left[1 + {1\over c^2} \left( \Pi - U  + {3p\over \rho} \right)  \right] r^2 = 4\pi \mathcal R r^2 \ .
    \label{pwmass}
\end{equation}
Using this mass to source the potential, we have
\begin{equation}
    {d U \over dr} = - {GM\over r^2} \ .
\end{equation}

The new mass $M$ is identical to the ADM mass (by definition equal to the gravitational mass) at 1pN order. To demonstrate this we may use, for example, equation (8.136) in \cite{PWbook} (see also \cite{1997PhRvD..55.6019S}), from which we have
\begin{equation}
    M_{\rm{ADM}} = \int \rho \left[ 1 + {1\over c^2} \left(  {5U \over 2} + \Pi \right) \right] d^3x = \int \rho^* \left[ 1 + {1\over c^2} \left(   \Pi - {U \over 2}  \right) \right] d^3x \ .
\end{equation}

We may also derive the result from full relativity.  The starting point would be the Tolman mass \cite{1987rtc..book.....T,2010arXiv1010.5557O}, defined as
\begin{multline}
    M_T = {1\over c^2} \int (\varepsilon+3p) \sqrt{-g}\, d^3x = \int  \rho \left[ 1 + {1\over c^2} \left(2U+ \Pi + {3p\over \rho} \right) \right] d^3x \\
     = \int  \rho \left[ 1 + {1\over c^2} \left({5U \over 2}+ \Pi \right) \right] d^3x = \int  \rho^\ast \left[ 1 + {1\over c^2} \left(\Pi - {U\over 2} \right) \right] d^3x \\ 
    =  \int  \rho \left[ 1 + {1\over c^2} \left(\Pi - {U\over 2} \right) \right] dV \ ,
    \label{gravmass}
\end{multline}
where we have used the virial relations from section~\ref{virel}. This argument shows that $M$ obtained from \eqref{pwmass} is, indeed, the 1pN version of the gravitational mass. Using the virial result, it is also easy see that \eqref{pwmass} leads to the same expression for $M$. The different results are consistent. The main insight at this stage is that the two masses $M_B$ and $M$ have clear physical meaning. 

We now get
\begin{multline}
       {dp \over dr} 
= - {G \rho^* \over r^2} \left\{ M_B +   {1\over c^2} \left[ \left( \Pi - 3U_N  + {p \over \rho^* }  \right)M_B + \mathcal N \right] \right\}  \\
= - \left( 1 - {2U \over c^2} \right) \left( 1 - {2p\over c^2} \right)  {G  M \mathcal R  \over r^2} \approx
- \left( 1 - {2U \over c^2} \right)   {G  M \mathcal R  \over r^2} + {2p\over c^2} {G \rho^* M_B \over r^2 } \\
= 
- \left( 1 - {2U \over c^2} \right)   {G  M \mathcal R  \over r^2} - {2p\over c^2} {dU \over dr} 
\label{BDSone}
\end{multline}
and if we define
\begin{equation}
    \mathcal P = \left( 1 + {2U \over c^2} \right) p
\end{equation}
we arrive at concise relation
\begin{equation}
    {d\mathcal P \over dr} = - { G  M \mathcal R \over r^2}
    \label{calP}
\end{equation}
Results obtained from \eqref{BDSone} are provided in figure~\ref{pwadjust}. This model does not seem to be an improvement on the versions considered in figure~\ref{PWfig}.

Finally, working with $M$ we may also use
\begin{equation}
    {dp\over dr} \approx - {GM \rho \over r^2} \left[ 1 + {1\over c^2} \left( \Pi + {p\over \rho} \right)\right] =  - {GM\over r^2 c^2}  (p+\varepsilon) \ .
    \label{BDStwo}
\end{equation}
Mass-radius results obtained from this model are provided in  figure~\ref{pwadjust}. This model performs rather well compared to the alternatives we have considered. As it does not suffer from the maximum density issue, this is an attractive (and comparatively simple) formulation of the pN problem.

Finally, we may ask if it is possible to take the notion of different masses one steps further. We have shown that one of the problematic features of the pN schemes relates to the a priori unknown central value of the gravitational potential. Given this, it is natural to ask if it is possible to  decouple the potential from the  equation for the pressure (as in the TOV equations). It turns out that we can, indeed, do this and the argument ultimately leads to us back to \eqref{wagmal}. As we have already considered this model, the step-by-step derivation is mainly relevant for completeness. The interested reader will find the derivation in the Appendix.

\section{Summary}
\label{sec:summary}

Neutron stars, with their strong gravitational fields, lie firmly in the relativistic regime. However, while a precise description of their structure and dynamics must be relativistic, there are a number of contexts where the modelling is technically challenging and where progress can be made by assuming weak gravity. With this in mind, we have explored the extent to which neutron stars can be described within the pN approximation. Indeed, working  to 1pN order, there is considerable freedom in how one formulates the equations (see, \textit{e.g.}, Refs.~\cite{1965ApJ...142.1488C,PWbook,1990MNRAS.242..289B}). A particularly important aspect relates to the freedom to introduce different mass functions in the stellar interior. We have shown that the choice of mass function may have a significant impact on the accuracy of   stellar models (in terms of mass and radius) and can, in some cases, lead to results that are fairly close to the fully relativistic result. In particular, our results show that neutron-star models based on \eqref{BDStwo}, which follows from a fairy simple prescription, are significantly more accurate than models built from \eqref{PWpressure}, which may be seen as the ``standard'' approach \cite{PWbook}. Here we demonstrated the result for a single matter equation of state, but we have confirmed it for a wide range of matter models.

As a step towards the pN discussion, we explored the use of isotropic coordinates. In contrast to the familiar TOV equations, the relativistic equations in such a coordinate system couple the two metric potentials. The problem also naturally involves two distinct mass functions (both of which  match the gravitational mass at the surface). We have devised a new strategy for solving  this coupled boundary-value problem, involving a scaling prescription that neatly circumvents the need for a root search. 

Motivated by the usual approach in the pN approximation, we  explored a truncated version of the isotropic TOV equations. Perhaps surprisingly, this uncovered a pathology in the isotropic formulation of the pN equations: there (typically) exists a \textit{maximum central density} beyond which one cannot find a numerical model that satisfies the boundary conditions. We showed that this problem occurs  because the pN expansion at these densities leads to a non-linear eigenvalue problem for the central value of the gravitational potential. Below the maximum density, there are two roots for a given central density, corresponding to a more and less massive star (only one of which is physical in the low-density limit).

Moving forward, one of the motivations for this work was to study the dynamical tides of neutron stars beyond Newtonian gravity (with a realistic matter model). While it is clear that the  pN strategy for building neutron-star models is far from robust (given the sizeable difference in results obtained from models that are---at least formally---consistent to the same pN order), the approach  does not suffer from the well-known issues that hamper the development of a fully relativistic description of dynamical tides. In particular, the set of stellar oscillation modes will remain complete at 1pN order, thus providing a potential avenue to develop more accurate descriptions of dynamical neutron-star tides using the mode-sum approach \cite{2020PhRvD.101h3001A}. Work in this direction will be relevant for future, sensitive gravitational-wave observations of compact binaries.

\section*{Acknowledgments}

NA and FG are grateful for support from STFC via grant numbers ST/R00045X/1 and ST/V000551/1. Parts of NA's contribution to this work were carried out at the Aspen Center for Physics, which is supported by National Science Foundation grant PHY-1607611. He also thanks the Simons Foundation for generous travel support.

\section*{Appendix: Decoupling the potential}

The purpose of this Appendix is to show that, if we set out to define a new mass function for which the gravitational potential is decoupled from the equation for the pressure, then we are naturally led back to equation~\eqref{wagmal}. 

The argument we need can be found in work by \citet{1974ApJ...189L..75W} and \citet{1981A&A....97L..12C,1983ApJ...275..867C} (see also the more recent discussion by \citet{2015PhRvD..92b4056G}) within the parameterised pN approach \cite{1993tegp.book.....W}. This strategy is commonly used to quantify the impact of deviations from general relativity and design tests of the theory. Our interest here is quite different. We want to explore different approximations to general relativity, rather than alternative theories. However, the scope does not matter. The steps we need are the same.

We basically want to ask if we can find a coordinate transformation that decouples the gravitational potential $U_N$ from the hydrostatic equations, taking as our starting point the (consistent) 1pN equation 
\begin{equation}
      \frac{dp}{dr} = - \frac{G \rho}{r^2} \left\{ M_N + \frac{1}{c^2} \left[ \left( \Pi + \frac{p}{\rho} \right) M_N + \mathcal N \right] \right\} \ ,
      \label{p1pN}
\end{equation}
where
\begin{equation}
    {d\mathcal{N} \over dr} =  4 \pi r^2 \rho \left( 2 U_N + \Pi + \frac{3p}{\rho} \right)  \ .
\end{equation}

Motivated by the transformation from isotropic to Schwarzschild coordinates from section~\ref{sec:isotropic} (noting that the TOV equations in Schwarzschild coordinates decouple from the metric potential $\nu$), we consider 
\begin{equation}
    \xi = r \left( 1 + \frac{1}{c^2} U_N \right) \ ,
\end{equation}
leading to
\begin{equation}
    \frac{d\xi}{dr} = 1 + \frac{1}{c^2} \left( U_N + r \frac{dU_N}{dr} \right) 
\end{equation}
and (keeping only terms up to 1pN order)
\begin{equation}
    \frac{dp}{d\xi} = \frac{dp}{dr} \frac{dr}{d\xi} = - \frac{G \rho}{\xi^2} \left\{ M_N + \frac{1}{c^2} \left[ \left( \Pi + \frac{p}{\rho} + U_N - \xi \frac{dU_N}{d\xi} \right) M_N + \mathcal{N} \right] \right\} \ .
\end{equation}
Similarly, we have 
\begin{equation}
    \frac{dM_N}{d\xi} = \frac{dM_N}{dr} \frac{dr}{d\xi} = 4 \pi \xi^2 \rho \left[ 1 - \frac{1}{c^2} \left( 3 U_N + \xi \frac{dU_N}{d\xi} \right) \right] 
\end{equation}
and 
\begin{equation}
    \frac{d\mathcal{N}}{d\xi} = \frac{d\mathcal{N}}{dr} \frac{dr}{d\xi} = 4 \pi \xi^2 \rho \left( 2 U_N + \Pi + 3 \frac{p}{\rho} \right) \ .
\end{equation}
Hence, we can write 
\begin{multline}
    \frac{dp}{d\xi} = - \frac{G \rho}{\xi^2} \Bigg\{ \int_0^\xi 4 \pi \bar{\xi}^2 \rho d\bar{\xi} + \frac{1}{c^2} \bigg[ \int_0^\xi 4 \pi \bar{\xi}^2 \rho \left( - U_N - \bar{\xi} \frac{dU_N}{d\bar{\xi}} + \Pi + 3 \frac{p}{\rho} \right) d\bar{\xi} \\
    + \left( \Pi + \frac{p}{\rho} + U_N - \xi \frac{dU_N}{d\xi} \right) \int_0^\xi 4 \pi \bar{\xi}^2 \rho d\bar{\xi} \bigg] \Bigg\} \ .
\end{multline}
It is now natural to define a mass function $\tilde{m}$ by 
\begin{equation}
    {d\tilde{m}  \over d\xi} =  4 \pi \xi^2 \rho  \ ,
\end{equation}
such that $M_N = \tilde{m} + \mathcal{O}(c^{-2})$. At this point, we need to do some integration by parts (making use of the relations from the virial theorem). First, we have
\begin{equation}
    I = \int_0^\xi 4 \pi \bar{\xi}^2 \rho \left( 3 \frac{p}{\rho} \right) d\bar{\xi} 
    = \left[ 4 \pi \bar{\xi}^3 p \right]_0^\xi - \int_0^\xi 4 \pi \bar{\xi}^3 \frac{dp}{d\bar{\xi}} d\bar{\xi} 
    = 4 \pi \xi^3 p - \int_0^\xi 4 \pi \bar{\xi}^3 \rho \frac{dU_N}{d\bar{\xi}} d\bar{\xi} \ .
\end{equation}
Therefore,  
\begin{equation}
    \frac{dp}{d\xi} = - \frac{G \rho}{\xi^2} \Bigg\{ \tilde{m} + \frac{1}{c^2} \bigg[ \int_0^\xi 4 \pi \bar{\xi}^2 \rho \left( - U_N - 2 \bar{\xi} \frac{dU_N}{d\bar{\xi}} + \Pi \right) d\bar{\xi} \\
    + 4 \pi \xi^3 p + \left( \Pi + \frac{p}{\rho} + U_N - \xi \frac{dU_N}{d\xi} \right) \tilde{m} \bigg] \Bigg\} \ .
\end{equation}
Next, we consider
\begin{equation}
    J = - \int_0^\xi 4 \pi \bar{\xi}^2 \rho U_N d\bar{\xi} = - \int_0^\xi \frac{d\tilde{m}}{d\bar{\xi}} U_N d\bar{\xi} = - \tilde{m} U_N + \int_0^\xi \tilde{m} \frac{dU_N}{d\bar{\xi}} d\bar{\xi} \ .
\end{equation}
We also need 
\begin{equation}
    K = \int_0^\xi \tilde{m} \frac{dU_N}{d\bar{\xi}} d\bar{\xi} 
    = - \int_0^\xi \frac{G \tilde{m}^2}{\bar{\xi}^2} d\bar{\xi}  = \left[ \frac{G \tilde{m}^2}{\bar{\xi}} \right]_0^\xi - \int_0^\xi \frac{2 G \tilde{m}}{\bar{\xi}} \frac{d\tilde{m}}{d\bar{\xi}} d\bar{\xi} \\
    = - \xi \frac{dU_N}{d\xi} \tilde{m} + \int_0^\xi 2 \bar{\xi} \frac{dU_N}{d\bar{\xi}} \frac{d\tilde{m}}{d\bar{\xi}} d\bar{\xi}  \ .
\end{equation}
Hence, 
\begin{equation}
    J = - \tilde{m} U_N - \xi \frac{dU_N}{d\xi} \tilde{m} + \int_0^\xi 2 \bar{\xi} \frac{dU_N}{d\bar{\xi}} \frac{d\tilde{m}}{d\bar{\xi}} d\bar{\xi}  \ .
\end{equation}
Finally, putting everything together, we arrive at
\begin{equation}
    \frac{dp}{d\xi} = - \frac{G \rho}{\xi^2} \left\{ \tilde{m} + \frac{1}{c^2} \left[ \int_0^\xi 4 \pi \bar{\xi}^2 \rho \Pi d\bar{\xi} + 4 \pi \xi^3 p + \left( \Pi + \frac{p}{\rho} + 2 \frac{G \tilde{m}}{\xi} \right) \tilde{m} \right] \right\} \ . 
\end{equation}
This is precisely what we wanted: an equation for hydrostatic equilibrium that does not explicitly involve the potential  $U_N$. The final expression is, however, not quite satisfactory as it requires the integral for the internal energy 
\begin{equation}
    E = \int_0^\xi 4 \pi \bar{\xi}^2 \rho \Pi d\bar{\xi}\ .
\end{equation}
The final step then is to introduce yet another mass function to remove this dependence on the energy from the equation for the pressure. The argument we need is discussed in detail by 
\citet{2015PhRvD..92b4056G}. We also need the gravitational potential energy, now given by  
\begin{equation}
    \Omega = - \int_0^\xi 4 \pi \bar{\xi}^2 \rho \frac{G \tilde{m}}{\bar{\xi}} d\bar{\xi} = - \int_0^\xi 4 \pi G \bar{\xi} \rho \tilde{m} d\bar{\xi}\ .
\end{equation}
In addition, we have terms that have dimensions of energy: $G \tilde{m}^2 / \xi$ and $4 \pi r^3 p$. Inspired by the parameterised pN strategy, a general mass function then takes the form
\begin{equation}
    \bar{m} = \tilde{m} + \frac{1}{c^2} \left( A E + B \Omega + C \frac{G \tilde{m}^2}{\xi} + D 4 \pi \xi^3 p \right).
\end{equation}
At leading order, this clearly reduces to the Newtonian mass $M_N$. It follows that
\begin{equation}
\begin{split}
    \frac{dp}{d\xi} = - \frac{G \rho}{\xi^2} \Bigg\{ \bar{m} + \frac{1}{c^2} \bigg[ (1 - A) E - B \Omega + (1 - D) 4 \pi \xi^3 p \\ 
    + \left( \Pi + \frac{p}{\rho} + (2 - C) \frac{G \bar{m}}{\xi} \right) \bar{m} \bigg] \Bigg\}  
\end{split}
\end{equation}
and 
\begin{equation}
    \frac{d\bar{m}}{d\xi} = 4 \pi \xi^2 \rho \left\{ 1 + \frac{1}{c^2} \left[ A \Pi + (2 C - B - D) \frac{G \bar{m}}{\xi} - C \frac{G \bar{m}^2}{4 \pi \xi^4 \rho} + 3 D \frac{p}{\rho} \right] \right\}  \ .
\end{equation}%
If we now choose $A = C = D = 1$, $B = 0$, we retain the expressions from \citet{1974ApJ...189L..75W}, i.e. the pressure is determined from \eqref{wagmal} which we already derived from the truncated TOV equation in section~\ref{tovtrunc}. This makes the argument somewhat ``circular'' but it is rewarding to see that the expressions can be derived from the 1pN relation \eqref{p1pN}.

\bibliography{biblio}

\begin{thebibliography}{25}%
\makeatletter
\providecommand \@ifxundefined [1]{%
 \@ifx{#1\undefined}
}%
\providecommand \@ifnum [1]{%
 \ifnum #1\expandafter \@firstoftwo
 \else \expandafter \@secondoftwo
 \fi
}%
\providecommand \@ifx [1]{%
 \ifx #1\expandafter \@firstoftwo
 \else \expandafter \@secondoftwo
 \fi
}%
\providecommand \natexlab [1]{#1}%
\providecommand \enquote  [1]{``#1''}%
\providecommand \bibnamefont  [1]{#1}%
\providecommand \bibfnamefont [1]{#1}%
\providecommand \citenamefont [1]{#1}%
\providecommand \href@noop [0]{\@secondoftwo}%
\providecommand \href [0]{\begingroup \@sanitize@url \@href}%
\providecommand \@href[1]{\@@startlink{#1}\@@href}%
\providecommand \@@href[1]{\endgroup#1\@@endlink}%
\providecommand \@sanitize@url [0]{\catcode `\\12\catcode `\$12\catcode
  `\&12\catcode `\#12\catcode `\^12\catcode `\_12\catcode `\%12\relax}%
\providecommand \@@startlink[1]{}%
\providecommand \@@endlink[0]{}%
\providecommand \url  [0]{\begingroup\@sanitize@url \@url }%
\providecommand \@url [1]{\endgroup\@href {#1}{\urlprefix }}%
\providecommand \urlprefix  [0]{URL }%
\providecommand \Eprint [0]{\href }%
\providecommand \doibase [0]{https://doi.org/}%
\providecommand \selectlanguage [0]{\@gobble}%
\providecommand \bibinfo  [0]{\@secondoftwo}%
\providecommand \bibfield  [0]{\@secondoftwo}%
\providecommand \translation [1]{[#1]}%
\providecommand \BibitemOpen [0]{}%
\providecommand \bibitemStop [0]{}%
\providecommand \bibitemNoStop [0]{.\EOS\space}%
\providecommand \EOS [0]{\spacefactor3000\relax}%
\providecommand \BibitemShut  [1]{\csname bibitem#1\endcsname}%
\let\auto@bib@innerbib\@empty
%</preamble>
\bibitem [{\citenamefont {{Andersson}}(2019)}]{NAbook}%
  \BibitemOpen
  \bibfield  {author} {\bibinfo {author} {\bibfnamefont {N.}~\bibnamefont
  {{Andersson}}},\ }\href
  {https://doi.org/10.1093/oso/9780198568032.001.0001/oso-9780198568032} {\emph
  {\bibinfo {title} {{Gravitational-Wave Astronomy: Exploring the Dark Side of
  the Universe}}}}\ (\bibinfo  {publisher} {Oxford University Press, Oxford},\
  \bibinfo {year} {2019})\BibitemShut {NoStop}%
\bibitem [{\citenamefont {{Abbott}}\ \emph {et~al.}(2017)\citenamefont
  {{Abbott}} \emph {et~al.}}]{2017PhRvL.119p1101A}%
  \BibitemOpen
  \bibfield  {author} {\bibinfo {author} {\bibfnamefont {B.~P.}\ \bibnamefont
  {{Abbott}}} \emph {et~al.},\ }\bibfield  {title} {\bibinfo {title}
  {{GW170817: Observation of Gravitational Waves from a Binary Neutron Star
  Inspiral}},\ }\href {https://doi.org/10.1103/PhysRevLett.119.161101}
  {\bibfield  {journal} {\bibinfo  {journal} {\prl}\ }\textbf {\bibinfo
  {volume} {119}},\ \bibinfo {eid} {161101} (\bibinfo {year}
  {2017})}\BibitemShut {NoStop}%
\bibitem [{\citenamefont {{Riles}}()}]{2022arXiv220606447R}%
  \BibitemOpen
  \bibfield  {author} {\bibinfo {author} {\bibfnamefont {K.}~\bibnamefont
  {{Riles}}},\ }\href@noop {} {\bibinfo {title} {{Searches for Continuous-Wave
  Gravitational Radiation}}},\ \Eprint {https://arxiv.org/abs/2206.06447}
  {arXiv:2206.06447 [astro-ph.HE]} \BibitemShut {NoStop}%
\bibitem [{\citenamefont {{Poisson}}\ and\ \citenamefont
  {{Will}}(2014)}]{PWbook}%
  \BibitemOpen
  \bibfield  {author} {\bibinfo {author} {\bibfnamefont {E.}~\bibnamefont
  {{Poisson}}}\ and\ \bibinfo {author} {\bibfnamefont {C.~M.}\ \bibnamefont
  {{Will}}},\ }\href {https://doi.org/10.1017/CBO9781139507486} {\emph
  {\bibinfo {title} {{Gravity}}}}\ (\bibinfo  {publisher} {Cambridge University
  Press, Cambridge},\ \bibinfo {year} {2014})\BibitemShut {NoStop}%
\bibitem [{\citenamefont {{Chatziioannou}}(2022)}]{2022PhRvD.105h4021C}%
  \BibitemOpen
  \bibfield  {author} {\bibinfo {author} {\bibfnamefont {K.}~\bibnamefont
  {{Chatziioannou}}},\ }\bibfield  {title} {\bibinfo {title} {{Uncertainty
  limits on neutron star radius measurements with gravitational waves}},\
  }\href {https://doi.org/10.1103/PhysRevD.105.084021} {\bibfield  {journal}
  {\bibinfo  {journal} {\prd}\ }\textbf {\bibinfo {volume} {105}},\ \bibinfo
  {eid} {084021} (\bibinfo {year} {2022})}\BibitemShut {NoStop}%
\bibitem [{\citenamefont {{Andersson}}\ and\ \citenamefont
  {{Pnigouras}}(2020)}]{2020PhRvD.101h3001A}%
  \BibitemOpen
  \bibfield  {author} {\bibinfo {author} {\bibfnamefont {N.}~\bibnamefont
  {{Andersson}}}\ and\ \bibinfo {author} {\bibfnamefont {P.}~\bibnamefont
  {{Pnigouras}}},\ }\bibfield  {title} {\bibinfo {title} {{Exploring the
  effective tidal deformability of neutron stars}},\ }\href
  {https://doi.org/10.1103/PhysRevD.101.083001} {\bibfield  {journal} {\bibinfo
   {journal} {\prd}\ }\textbf {\bibinfo {volume} {101}},\ \bibinfo {eid}
  {083001} (\bibinfo {year} {2020})}\BibitemShut {NoStop}%
\bibitem [{\citenamefont {{Friedman}}\ and\ \citenamefont
  {{Schutz}}(1978{\natexlab{a}})}]{1978ApJ...221..937F}%
  \BibitemOpen
  \bibfield  {author} {\bibinfo {author} {\bibfnamefont {J.~L.}\ \bibnamefont
  {{Friedman}}}\ and\ \bibinfo {author} {\bibfnamefont {B.~F.}\ \bibnamefont
  {{Schutz}}},\ }\bibfield  {title} {\bibinfo {title} {{Lagrangian perturbation
  theory of nonrelativistic fluids.}},\ }\href {https://doi.org/10.1086/156098}
  {\bibfield  {journal} {\bibinfo  {journal} {\apj}\ }\textbf {\bibinfo
  {volume} {221}},\ \bibinfo {pages} {937} (\bibinfo {year}
  {1978}{\natexlab{a}})}\BibitemShut {NoStop}%
\bibitem [{\citenamefont {{Friedman}}\ and\ \citenamefont
  {{Schutz}}(1978{\natexlab{b}})}]{1978ApJ...222..281F}%
  \BibitemOpen
  \bibfield  {author} {\bibinfo {author} {\bibfnamefont {J.~L.}\ \bibnamefont
  {{Friedman}}}\ and\ \bibinfo {author} {\bibfnamefont {B.~F.}\ \bibnamefont
  {{Schutz}}},\ }\bibfield  {title} {\bibinfo {title} {{Secular instability of
  rotating Newtonian stars.}},\ }\href {https://doi.org/10.1086/156143}
  {\bibfield  {journal} {\bibinfo  {journal} {\apj}\ }\textbf {\bibinfo
  {volume} {222}},\ \bibinfo {pages} {281} (\bibinfo {year}
  {1978}{\natexlab{b}})}\BibitemShut {NoStop}%
\bibitem [{\citenamefont {{Gundlach}}\ \emph {et~al.}(1994)\citenamefont
  {{Gundlach}}, \citenamefont {{Price}},\ and\ \citenamefont
  {{Pullin}}}]{1994PhRvD..49..883G}%
  \BibitemOpen
  \bibfield  {author} {\bibinfo {author} {\bibfnamefont {C.}~\bibnamefont
  {{Gundlach}}}, \bibinfo {author} {\bibfnamefont {R.~H.}\ \bibnamefont
  {{Price}}},\ and\ \bibinfo {author} {\bibfnamefont {J.}~\bibnamefont
  {{Pullin}}},\ }\bibfield  {title} {\bibinfo {title} {{Late-time behavior of
  stellar collapse and explosions. I. Linearized perturbations}},\ }\href
  {https://doi.org/10.1103/PhysRevD.49.883} {\bibfield  {journal} {\bibinfo
  {journal} {\prd}\ }\textbf {\bibinfo {volume} {49}},\ \bibinfo {pages} {883}
  (\bibinfo {year} {1994})}\BibitemShut {NoStop}%
\bibitem [{\citenamefont {{Nollert}}(1999)}]{1999CQGra..16R.159N}%
  \BibitemOpen
  \bibfield  {author} {\bibinfo {author} {\bibfnamefont {H.-P.}\ \bibnamefont
  {{Nollert}}},\ }\bibfield  {title} {\bibinfo {title} {{TOPICAL REVIEW:
  Quasinormal modes: the characteristic `sound' of black holes and neutron
  stars}},\ }\href {https://doi.org/10.1088/0264-9381/16/12/201} {\bibfield
  {journal} {\bibinfo  {journal} {Classical Quant. Grav.}\ }\textbf {\bibinfo
  {volume} {16}},\ \bibinfo {pages} {R159} (\bibinfo {year}
  {1999})}\BibitemShut {NoStop}%
\bibitem [{\citenamefont {{Chandrasekhar}}(1965)}]{1965ApJ...142.1488C}%
  \BibitemOpen
  \bibfield  {author} {\bibinfo {author} {\bibfnamefont {S.}~\bibnamefont
  {{Chandrasekhar}}},\ }\bibfield  {title} {\bibinfo {title} {{The
  Post-Newtonian Equations of Hydrodynamics in General Relativity.}},\ }\href
  {https://doi.org/10.1086/148432} {\bibfield  {journal} {\bibinfo  {journal}
  {\apj}\ }\textbf {\bibinfo {volume} {142}},\ \bibinfo {pages} {1488}
  (\bibinfo {year} {1965})}\BibitemShut {NoStop}%
\bibitem [{\citenamefont {{Glampedakis}}\ \emph {et~al.}(2015)\citenamefont
  {{Glampedakis}}, \citenamefont {{Pappas}}, \citenamefont {{Silva}},\ and\
  \citenamefont {{Berti}}}]{2015PhRvD..92b4056G}%
  \BibitemOpen
  \bibfield  {author} {\bibinfo {author} {\bibfnamefont {K.}~\bibnamefont
  {{Glampedakis}}}, \bibinfo {author} {\bibfnamefont {G.}~\bibnamefont
  {{Pappas}}}, \bibinfo {author} {\bibfnamefont {H.~O.}\ \bibnamefont
  {{Silva}}},\ and\ \bibinfo {author} {\bibfnamefont {E.}~\bibnamefont
  {{Berti}}},\ }\bibfield  {title} {\bibinfo {title}
  {{Post-Tolman-Oppenheimer-Volkoff formalism for relativistic stars}},\ }\href
  {https://doi.org/10.1103/PhysRevD.92.024056} {\bibfield  {journal} {\bibinfo
  {journal} {\prd}\ }\textbf {\bibinfo {volume} {92}},\ \bibinfo {eid} {024056}
  (\bibinfo {year} {2015})}\BibitemShut {NoStop}%
\bibitem [{\citenamefont {{Blanchet}}\ \emph {et~al.}(1990)\citenamefont
  {{Blanchet}}, \citenamefont {{Damour}},\ and\ \citenamefont
  {{Schaefer}}}]{1990MNRAS.242..289B}%
  \BibitemOpen
  \bibfield  {author} {\bibinfo {author} {\bibfnamefont {L.}~\bibnamefont
  {{Blanchet}}}, \bibinfo {author} {\bibfnamefont {T.}~\bibnamefont
  {{Damour}}},\ and\ \bibinfo {author} {\bibfnamefont {G.}~\bibnamefont
  {{Schaefer}}},\ }\bibfield  {title} {\bibinfo {title} {{Post-Newtonian
  hydrodynamics and post-Newtonian gravitational wave generation for numerical
  relativity}},\ }\href {https://doi.org/10.1093/mnras/242.3.289} {\bibfield
  {journal} {\bibinfo  {journal} {Mon. Not. Astron. Soc.}\ }\textbf {\bibinfo
  {volume} {242}},\ \bibinfo {pages} {289} (\bibinfo {year}
  {1990})}\BibitemShut {NoStop}%
\bibitem [{\citenamefont {{Misner}}\ \emph {et~al.}(2017)\citenamefont
  {{Misner}}, \citenamefont {{Thorne}},\ and\ \citenamefont
  {{Wheeler}}}]{2017grav.book.....M}%
  \BibitemOpen
  \bibfield  {author} {\bibinfo {author} {\bibfnamefont {C.~W.}\ \bibnamefont
  {{Misner}}}, \bibinfo {author} {\bibfnamefont {K.~S.}\ \bibnamefont
  {{Thorne}}},\ and\ \bibinfo {author} {\bibfnamefont {J.~A.}\ \bibnamefont
  {{Wheeler}}},\ }\href@noop {} {\emph {\bibinfo {title} {{Gravitation}}}}\
  (\bibinfo  {publisher} {Princeton University Press, Princeton},\ \bibinfo
  {year} {2017})\BibitemShut {NoStop}%
\bibitem [{\citenamefont {{Shinkai}}(1999)}]{1999PhRvD..60f7504S}%
  \BibitemOpen
  \bibfield  {author} {\bibinfo {author} {\bibfnamefont {H.-A.}\ \bibnamefont
  {{Shinkai}}},\ }\bibfield  {title} {\bibinfo {title} {{Truncated
  post-Newtonian neutron star model}},\ }\href
  {https://doi.org/10.1103/PhysRevD.60.067504} {\bibfield  {journal} {\bibinfo
  {journal} {\prd}\ }\textbf {\bibinfo {volume} {60}},\ \bibinfo {eid} {067504}
  (\bibinfo {year} {1999})}\BibitemShut {NoStop}%
\bibitem [{\citenamefont {{Goriely}}\ \emph {et~al.}(2013)\citenamefont
  {{Goriely}}, \citenamefont {{Chamel}},\ and\ \citenamefont
  {{Pearson}}}]{2013PhRvC..88b4308G}%
  \BibitemOpen
  \bibfield  {author} {\bibinfo {author} {\bibfnamefont {S.}~\bibnamefont
  {{Goriely}}}, \bibinfo {author} {\bibfnamefont {N.}~\bibnamefont
  {{Chamel}}},\ and\ \bibinfo {author} {\bibfnamefont {J.~M.}\ \bibnamefont
  {{Pearson}}},\ }\bibfield  {title} {\bibinfo {title} {{Further explorations
  of Skyrme-Hartree-Fock-Bogoliubov mass formulas. XIII. The 2012 atomic mass
  evaluation and the symmetry coefficient}},\ }\href
  {https://doi.org/10.1103/PhysRevC.88.024308} {\bibfield  {journal} {\bibinfo
  {journal} {\prc}\ }\textbf {\bibinfo {volume} {88}},\ \bibinfo {eid} {024308}
  (\bibinfo {year} {2013})}\BibitemShut {NoStop}%
\bibitem [{\citenamefont {{Wagoner}}\ and\ \citenamefont
  {{Malone}}(1974)}]{1974ApJ...189L..75W}%
  \BibitemOpen
  \bibfield  {author} {\bibinfo {author} {\bibfnamefont {R.~V.}\ \bibnamefont
  {{Wagoner}}}\ and\ \bibinfo {author} {\bibfnamefont {R.~C.}\ \bibnamefont
  {{Malone}}},\ }\bibfield  {title} {\bibinfo {title} {{Post-Newtonian Neutron
  Stars}},\ }\href {https://doi.org/10.1086/181468} {\bibfield  {journal}
  {\bibinfo  {journal} {\apj}\ }\textbf {\bibinfo {volume} {189}},\ \bibinfo
  {pages} {L75} (\bibinfo {year} {1974})}\BibitemShut {NoStop}%
\bibitem [{\citenamefont {{Tsokaros}}\ \emph {et~al.}(2015)\citenamefont
  {{Tsokaros}}, \citenamefont {{Uryu}},\ and\ \citenamefont
  {{Rezzolla}}}]{2015PhRvD..91j4030T}%
  \BibitemOpen
  \bibfield  {author} {\bibinfo {author} {\bibfnamefont {A.}~\bibnamefont
  {{Tsokaros}}}, \bibinfo {author} {\bibfnamefont {K.}~\bibnamefont {{Uryu}}},\
  and\ \bibinfo {author} {\bibfnamefont {L.}~\bibnamefont {{Rezzolla}}},\
  }\bibfield  {title} {\bibinfo {title} {{New code for quasiequilibrium initial
  data of binary neutron stars: Corotating, irrotational, and slowly spinning
  systems}},\ }\href {https://doi.org/10.1103/PhysRevD.91.104030} {\bibfield
  {journal} {\bibinfo  {journal} {\prd}\ }\textbf {\bibinfo {volume} {91}},\
  \bibinfo {eid} {104030} (\bibinfo {year} {2015})}\BibitemShut {NoStop}%
\bibitem [{\citenamefont {{Raaijmakers}}\ \emph {et~al.}(2021)\citenamefont
  {{Raaijmakers}}, \citenamefont {{Greif}}, \citenamefont {{Hebeler}},
  \citenamefont {{Hinderer}}, \citenamefont {{Nissanke}}, \citenamefont
  {{Schwenk}}, \citenamefont {{Riley}}, \citenamefont {{Watts}}, \citenamefont
  {{Lattimer}},\ and\ \citenamefont {{Ho}}}]{2021ApJ...918L..29R}%
  \BibitemOpen
  \bibfield  {author} {\bibinfo {author} {\bibfnamefont {G.}~\bibnamefont
  {{Raaijmakers}}}, \bibinfo {author} {\bibfnamefont {S.~K.}\ \bibnamefont
  {{Greif}}}, \bibinfo {author} {\bibfnamefont {K.}~\bibnamefont {{Hebeler}}},
  \bibinfo {author} {\bibfnamefont {T.}~\bibnamefont {{Hinderer}}}, \bibinfo
  {author} {\bibfnamefont {S.}~\bibnamefont {{Nissanke}}}, \bibinfo {author}
  {\bibfnamefont {A.}~\bibnamefont {{Schwenk}}}, \bibinfo {author}
  {\bibfnamefont {T.~E.}\ \bibnamefont {{Riley}}}, \bibinfo {author}
  {\bibfnamefont {A.~L.}\ \bibnamefont {{Watts}}}, \bibinfo {author}
  {\bibfnamefont {J.~M.}\ \bibnamefont {{Lattimer}}},\ and\ \bibinfo {author}
  {\bibfnamefont {W.~C.~G.}\ \bibnamefont {{Ho}}},\ }\bibfield  {title}
  {\bibinfo {title} {{Constraints on the Dense Matter Equation of State and
  Neutron Star Properties from NICER's Mass-Radius Estimate of PSR J0740+6620
  and Multimessenger Observations}},\ }\href
  {https://doi.org/10.3847/2041-8213/ac089a} {\bibfield  {journal} {\bibinfo
  {journal} {\apj}\ }\textbf {\bibinfo {volume} {918}},\ \bibinfo {eid} {L29}
  (\bibinfo {year} {2021})}\BibitemShut {NoStop}%
\bibitem [{\citenamefont {{Will}}(1993)}]{1993tegp.book.....W}%
  \BibitemOpen
  \bibfield  {author} {\bibinfo {author} {\bibfnamefont {C.~M.}\ \bibnamefont
  {{Will}}},\ }\href {https://doi.org/10.1017/9781316338612} {\emph {\bibinfo
  {title} {{Theory and Experiment in Gravitational Physics}}}}\ (\bibinfo
  {publisher} {Cambridge University Press, Cambridge},\ \bibinfo {year}
  {1993})\BibitemShut {NoStop}%
\bibitem [{\citenamefont {{Shibata}}(1997)}]{1997PhRvD..55.6019S}%
  \BibitemOpen
  \bibfield  {author} {\bibinfo {author} {\bibfnamefont {M.}~\bibnamefont
  {{Shibata}}},\ }\bibfield  {title} {\bibinfo {title} {{Numerical study of
  synchronized binary neutron stars in the post-Newtonian approximation of
  general relativity}},\ }\href {https://doi.org/10.1103/PhysRevD.55.6019}
  {\bibfield  {journal} {\bibinfo  {journal} {\prd}\ }\textbf {\bibinfo
  {volume} {55}},\ \bibinfo {pages} {6019} (\bibinfo {year}
  {1997})}\BibitemShut {NoStop}%
\bibitem [{\citenamefont {{Tolman}}(1987)}]{1987rtc..book.....T}%
  \BibitemOpen
  \bibfield  {author} {\bibinfo {author} {\bibfnamefont {R.~C.}\ \bibnamefont
  {{Tolman}}},\ }\href@noop {} {\emph {\bibinfo {title} {{Relativity,
  thermodynamics and cosmology}}}}\ (\bibinfo  {publisher} {Dover, Mineola},\
  \bibinfo {year} {1987})\BibitemShut {NoStop}%
\bibitem [{\citenamefont {{Ohanian}}()}]{2010arXiv1010.5557O}%
  \BibitemOpen
  \bibfield  {author} {\bibinfo {author} {\bibfnamefont {H.~C.}\ \bibnamefont
  {{Ohanian}}},\ }\href@noop {} {\bibinfo {title} {{The Energy-Momentum Tensor
  in General Relativity and in Alternative Theories of Gravitation, and the
  Gravitational vs. Inertial Mass}}},\ \Eprint
  {https://arxiv.org/abs/1010.5557} {arXiv:1010.5557 [gr-qc]} \BibitemShut
  {NoStop}%
\bibitem [{\citenamefont {{Ciufolini}}\ and\ \citenamefont
  {{Ruffini}}(1981)}]{1981A&A....97L..12C}%
  \BibitemOpen
  \bibfield  {author} {\bibinfo {author} {\bibfnamefont {I.}~\bibnamefont
  {{Ciufolini}}}\ and\ \bibinfo {author} {\bibfnamefont {R.}~\bibnamefont
  {{Ruffini}}},\ }\bibfield  {title} {\bibinfo {title} {{On the value of the
  masses of neutron stars in the parameterized post Newtonian formalism}},\
  }\href@noop {} {\bibfield  {journal} {\bibinfo  {journal} {Astron.
  Astrophys.}\ }\textbf {\bibinfo {volume} {97}},\ \bibinfo {pages} {L12}
  (\bibinfo {year} {1981})}\BibitemShut {NoStop}%
\bibitem [{\citenamefont {{Ciufolini}}\ and\ \citenamefont
  {{Ruffini}}(1983)}]{1983ApJ...275..867C}%
  \BibitemOpen
  \bibfield  {author} {\bibinfo {author} {\bibfnamefont {I.}~\bibnamefont
  {{Ciufolini}}}\ and\ \bibinfo {author} {\bibfnamefont {R.}~\bibnamefont
  {{Ruffini}}},\ }\bibfield  {title} {\bibinfo {title} {{Equilibrium
  configurations of neutron stars and the parametrized post-Newtonian metric
  theories of gravitation}},\ }\href {https://doi.org/10.1086/161580}
  {\bibfield  {journal} {\bibinfo  {journal} {\apj}\ }\textbf {\bibinfo
  {volume} {275}},\ \bibinfo {pages} {867} (\bibinfo {year}
  {1983})}\BibitemShut {NoStop}%
\end{thebibliography}%

\end{document}